\documentclass[fleqn,usenatbib]{mnras}

\usepackage{newtxtext,newtxmath}

\usepackage[T1]{fontenc}
\usepackage{ae,aecompl}

\usepackage{graphicx}	
\usepackage{amsmath}	
\usepackage{amssymb}	
\usepackage{hyperref}
\usepackage{soul}

\newcommand{\gadget}[1]{\texttt{GADGET-2}#1} 
\newcommand{\galic}[1]{\texttt{GalIC}#1} 
\newcommand{\simspin}[1]{\texttt{SimSpin}#1} 

\title[The limitations of $\lambda_R$]{A numerical twist on the spin parameter, $\lambda_R$}

\author[K.E. Harborne et al.]{K.E. Harborne,$^{1,2,3}$\thanks{E-mail: katherine.harborne@icrar.org}
C. Power,$^{1,2,3}$
A.S.G. Robotham,$^{1,2}$
L. Cortese,$^{1,2}$ 
\newauthor and D.S. Taranu$^{1,3}$ \\
$^{1}$International Centre for Radio Astronomy (ICRAR), M468, The University of Western Australia, 35 Stirling Highway, \\ 
Crawley, WA 6009, Australia\\
$^{2}$ARC Centre of Excellence for All Sky Astrophysics in 3 Dimensions (ASTRO 3D) \\
$^{3}$ARC Centre of Excellence for All-sky Astrophysics (CAASTRO)
}

\date{Accepted XXX. Received YYY; in original form ZZZ}

\pubyear{2017}

\begin{document}
\label{firstpage}
\pagerange{\pageref{firstpage}--\pageref{lastpage}}
\maketitle

\begin{abstract}
A primary goal of integral field spectroscopic (IFS) surveys is to provide a statistical census of galaxies classified by their internal kinematics. As a result, the observational spin parameter, $\lambda_R$, has become one of the most popular methods of quantifying the relative importance of velocity dispersion and rotation in supporting a galaxy's inner structure. The goal of this paper is to examine the relationship between the observationally deduced $\lambda_R$ and one of the most commonly used theoretical spin parameters in the literature, the \citet{2001_bullock} $\lambda'$. Using a set of $N$-body realisations of galaxies from which we construct mock IFS observations, we measure $\lambda_R$ as an observer would, incorporating the effects of beam smearing and seeing conditions. Assuming parameters typical of current IFS surveys, we confirm that there are strong positive correlations between $\lambda_R$ and measurement radius, and strong negative correlations between $\lambda_R$ and size of the PSF, for late-type galaxies; these  biases can be reduced using a recently proposed empirical correction. Once observational biases are corrected for, we find that $\lambda_R$ provides a good approximation to $\sim \sqrt{3}/2 \; \lambda'(\rm R_{\rm eff})$, where $\lambda'$ is evaluated for the galactic stellar component within 1 R$_{\rm eff}$.
\end{abstract}

\begin{keywords}
Galaxy: evolution -- galaxies: kinematics and dynamics 
\end{keywords}



\section{Introduction}

Our aim, as extra-galactic astronomers, is to understand the formation and evolutionary mechanisms that lead to the variety of galaxies that we observe across the Universe. The first step in approaching any development in science is to categorise the objects under scrutiny. Beginning with \cite{1926_hubble}, \cite{1959_deVaucouleurs}, \cite{1975_sandage} and many others, a visual classification of galaxy morphology has been established with which we can investigate patterns and generate hypotheses about the evolutionary paths that these objects have followed. While it is useful to group galaxies by their appearance, all features need to be explained by some form of physical process. With this in mind, it is necessary to further constrain galaxy classifications, and hence models of formation, by folding in our growing knowledge of stellar kinematics \citep{2014_conselice}. 

Traditionally, a galaxy's kinematics have been characterised by taking measurements in two separate apertures - in a long-slit aligned with the galaxy's major axis to find the line-of-sight (LOS) velocity, $V$; and in a central aperture to calculate its LOS velocity dispersion, $\sigma$ \citep{1977_illingworth, 1983_davies, 1988_bender}. Combining these measurements into the ratio, $V/\sigma$, then provides a means to quantify the relative importance of ordered versus unordered motions in the galaxy. However, galaxies can have complex kinematic structures that are not easily characterised by $V/\sigma$, as demonstrated by the results of \citet{2002_deZeeuw} and \citet{2004_emsellem}, who used the integral field Spectrographic Areal Unit for Research on Optical Nebulae (SAURON) survey to show E/S0 galaxies have a diverse range kinematic structures; for example, galaxies with kinematically decoupled cores and galaxies that exhibit regular rotation both occupy similar regions of $V/\sigma$ parameter space \citep{2007_emsellem}. 

A luminosity-weighted measure of galaxy rotation has since become fashionable. In
\citeyear{2007_emsellem}, \citeauthor{2007_emsellem} presented SAURON galaxies in a parameter space described by an observational spin parameter, $\lambda_R$, and ellipticity, $\epsilon$. $\lambda_R$ utilises the radial distribution of kinematics provided by integral field spectroscopy (IFS), defined by,
\begin{equation}
\lambda_R \equiv \frac{\left<R |V|\right>}{\left<R \sqrt{V^2 + \sigma^2}\right>} \;,
\label{eq:lambda1}
\end{equation}
where $R$ is the circularised radial position, $V$ is the LOS velocity, and $\sigma$ is the LOS velocity dispersion taken as luminosity-weighted averages denoted by the angular brackets. With their internal kinematics defined in this way, galaxies split into two distinct groups that mirror the morphology division between ellipticals and spirals: slow rotators (SR) and fast rotators (FR). ATLAS$^{\text{3D}}$ \citep{2011_cappellari_b} gave the first statistical sample of galaxies within which these spin properties could be examined. 

With the emergence of large IFS surveys (SAMI, Sydney-AAO Multi-object Integral field spectrometer, \citealt{2011_croom}; CALIFA, Calar Alto Legacy Integral Field Area, \citealt{2012_sanchez}; and MaNGA, Mapping Nearby Galaxies at Apache point observatory, \citealt{2015_bundy}), it is now possible to map gas and stellar motions, and to measure kinematic quantities such as $\lambda_R$, in much greater statistical samples of galaxies. These datasets are being used to study the relationship between $\lambda_R$, galaxy mass, and environment \citep{2017_veale, 2017b_sande, 2017_brough, 2018_smethurst}, and, as a consequence, are providing unique insights into the assembly history of galaxies \citep{2013_foster, 2014_arnold, 2015_fogarty, 2016_cortese}. This work is complemented naturally by the emergence of galaxy formation simulations of cosmological volumes, such as the {\small{Illustris}} simulation \citep{2014_vogelsberger} and {\small{EAGLE}} simulations \citep{2015_schaye}. These simulations, which track galaxy growth over cosmic time and span a range of galaxy masses, environments, and assembly histories, provide a powerful tool to study the physical origin of galactic angular momentum \citep{2015_teklu, 2015_genel, 2015_pedrossa, 2016_zavala, 2018_lagos}, and to understand the astrophysical implications of the observational data, including $V/\sigma$ and $\lambda_R$.
\bigskip 

The utility of numerical simulations to interpret observational data relies upon our ability to compare observational and simulated data sets in as similar a manner as is possible. It is worth noting that the form of $\lambda_R$, originally proposed by \cite{2007_emsellem}, was constructed to provide an observational proxy for the total amount of angular momentum per unit mass, which is a natural analogue of the theoretical spin parameter widely used in numerical simulations. 

Two definitions of the theoretical spin parameter are commonplace in the literature. The first is the \citet{1969_peebles} $\lambda$; here,
\begin{equation}
\label{eq:spin1}
\lambda = \frac{J|E|^{1/2}}{G M^{5/2}} \,;
\end{equation}
where $\textbf{J}$ is angular momentum, $E$ is total energy, $M$ is the virial mass and $G$ is the gravitational constant. Eq.~\ref{eq:spin1} quantifies the relative importance of rotational and dispersion support to maintaining the entire gravitationally bound structure in equilibrium, and it is often used to measure specific angular momentum \citep[e.g.][]{1980_fall,1998_mo,2008_knebe}; indeed, \citet{2007_emsellem} showed how $\lambda$ and $\lambda_R$ connect to one another. However, this form has an explicit dependence on the total (i.e. kinetic plus gravitational) energy of the system and so special care must be taken when calculating the gravitational potential energy\footnote{This is because the gravitational potential depends on the mass distribution at all radii, not just interior to a given radius. A surface pressure correction can account for this - as used previously in cosmological $N$-body simulations \citep[e.g.][]{2006_shaw, 2012_power} - but the size of this correction tends increase with decreasing halo-centric radius decreases, and is likely to be large on the scale of galaxies.}, and in its standard form it is only applicable when considering all matter out to some encompassing radius \citep{2015_teklu}, usually the virial radius.

The second is the \citet{2001_bullock} $\lambda'$, given by,
\begin{equation}
\lambda' = \frac{J}{\sqrt{2} M V_{c} R} \;,
\label{eq:spin2}
\end{equation}
\noindent where $\textbf{J}$, $M$ and $V_c$ are the angular momentum, mass and circular velocity all measured within radius $R$. This form is particularly attractive because it depends only on material within $R$ and can be calculated for an individual component, $k$,
\begin{equation}
\lambda'_{k} = \frac{j_{k}}{\sqrt{2} V_{c} R} \;,
\label{eq:spin3}
\end{equation}
\noindent where the angular momentum is replaced with the specific angular momentum, $j_{k} = J_{k}/M_{k}$. In contrast to the Peebles form, Eq.~\ref{eq:spin2} and~\ref{eq:spin3} give us the ability to examine the radial distribution of the spin parameter straightforwardly.

When evaluated for all components within the virial radius, $\lambda \approx \lambda'$. Hence, when quoting the kinematic properties of a galaxy, typically $\lambda'$ is defined at the virial radius, $r_{vir}$. In contrast, $\lambda_R$ is usually measured within an effective radius, $R_{\rm eff}$, of the galaxy, which is much smaller than $r_{vir}$. Although \cite{2007_emsellem} provides an approximate conversion between $\lambda$ and $\lambda_R$, it is important to assess how well $\lambda_R$ deduced from observational data and $\lambda$/$\lambda'$ derived from numerical simulations relate to one another in detail.

Previous work has shown that $\lambda_R$ should be a robust tracer of intrinsic angular momentum \citep[e.g.][who analysed $N$-body simulated merger remnants]{2009_jesseit}.  The goal of this paper is to characterise the relationship between $\lambda_R$, as it may be observationally deduced, and $\lambda'$, as it is measured in a numerical simulation, for a set of $N$-body realisations of galaxies with varying bulge-to-total mass (B/T) ratios. By constructing mock IFS observations of our data, we can measure $\lambda_R$ as an observer would, incorporating the effects of beam smearing, seeing, and measurement radius. This allows us to assess how well we can correct the measured $\lambda_R$ to remove observational bias, and its relationship to the intrinsic $\lambda'$ of the galaxy.

Understanding these observational biases is crucial if we wish to compare observations and predictions from simulations. Spatial blurring caused by the Earth's atmosphere introduces uncertainties into kinematic measurements, such that velocity gradients are smoothed, LOS velocity dispersions are reduced, and measurements of properties like $\lambda_R$ become biased. Previous studies have sought to quantify or correct this bias \citep[e.g.][]{2013_deugenio, 2017b_sande, 2017a_sande, 2018_greene, 2018_graham}, and we investigate how well such corrections perform. In particular, we focus on the recent work \cite{2018_graham}, who used Jeans Anisotropic Modelling (JAM) to derive a general analytic correction to $\lambda_R$ for regularly rotating galaxies of different types, based on the ratio of the width of the point spread function ($\sigma_{\rm PSF}$) to the effective radius of the galaxy ($R_{\rm eff}$); we investigate how well it performs when applied to 3D dynamical galaxy models.

\bigskip

In the remaining sections of this paper, we describe the simulations and the
method used to construct our synthetic observations (\S~\ref{sec:method}). In
\S~\ref{sec:results}, we present the results of our investigations and go on
to discuss the outcome of these with respect to observational and definition
bias in section \ref{sec:discuss}. A detailed explanation of the galaxy models
is given in appendix A and a mathematical derivation of $\lambda_R$ from
$\lambda'$ is given in appendix B. Throughout, we assume a Lambda-cold dark
matter ($\Lambda CDM$) cosmology with
$\Omega_m = 0.308, \Omega_{\Lambda} = 0.692$ and $H_0 = 67.8$.

\begin{figure*}
\centering
\includegraphics[width=0.94\textwidth]{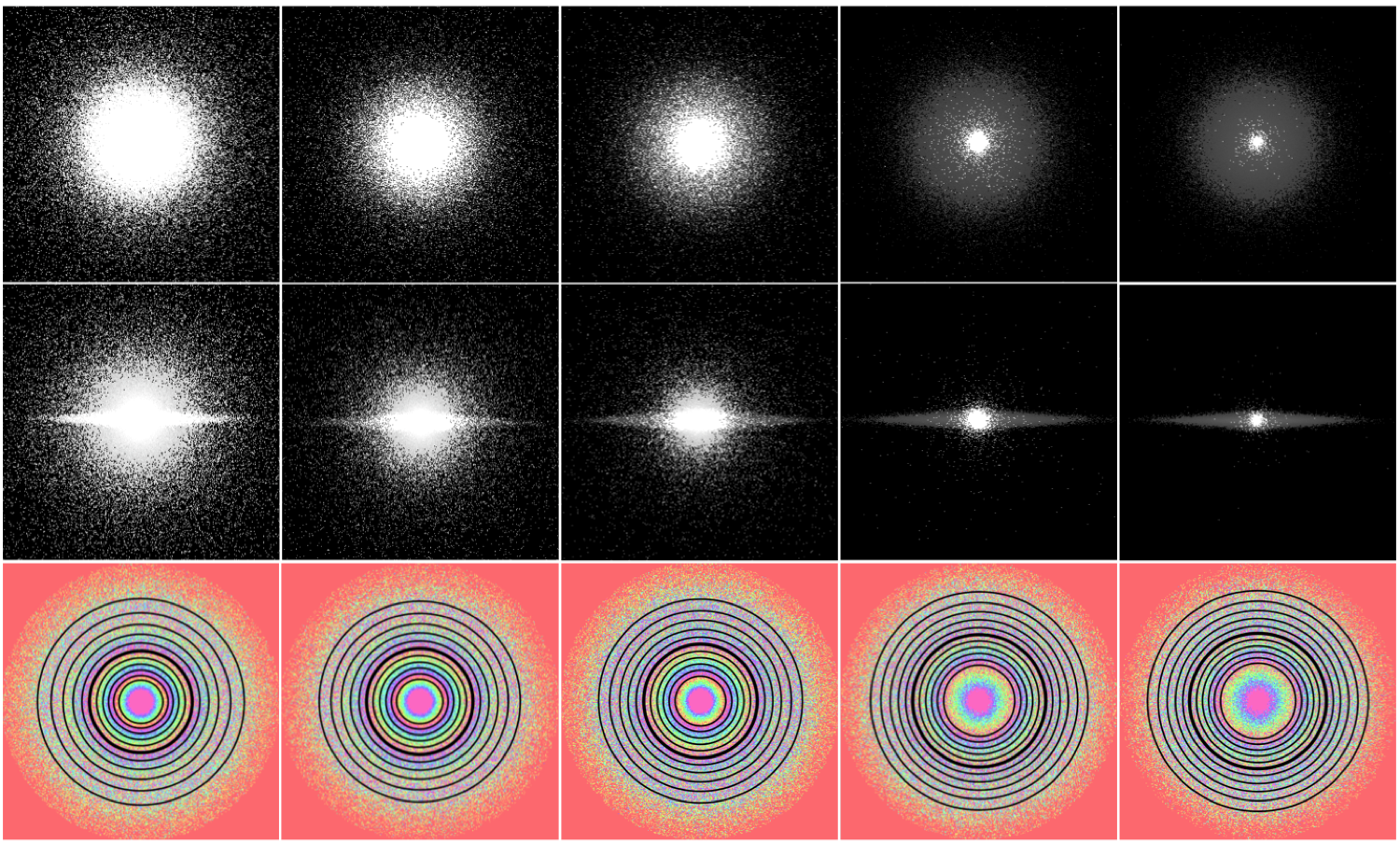}
\caption{The mock catalogue of galaxies investigated in this work. From left to right, we show the S0, Sa, Sb, Sc and Sd models. The top row shows each galaxy inclined to 0$^{\circ}$ (i.e. face-on) and the middle row demonstrates the same particle distribution projected at 90$^{\circ}$ (i.e. edge-on). These top two rows demonstrate the particles contained within a radius of 50kpc. The bottom row shows the isophotal images of each galaxy within the SAMI field of view inclined face on at 0$^{\circ}$. The overlaid ellipses contain 0.25-0.75 of the total flux at increments of 0.05. The $R_{\text{eff}}$ for each galaxy is shown by the bold black line.} 
\label{fig:cat}
\end{figure*}

\section{Method}
\label{sec:method}

We have constructed a catalogue of five isolated, stable $N$-body realisations of galaxies across a range of B/T ratios in order to explore the relationship between the observed $\lambda_R$ and intrinsic spin, $\lambda'$. The simple models in this catalogue are shown in Fig. \ref{fig:cat}. The construction details of these simulations are described in section \ref{sec:setup}. For each of the galaxy models, we make a series of synthetic observations using the R package \simspin \footnote{This code can be found in a github repository at \url{https://github.com/kateharborne/SimSpin}. Worked examples can be found at \url{https://rpubs.com/kateharborne}.}. This code allows the user to generate a kinematic data cube based on simulated particle positions and velocities, vary the level of spatial blurring and measure $\lambda_R$ within a specified radius. The method by which we make these mock observations is explained in section \ref{sec:observations}.

\subsection{The simulations}
\label{sec:setup}

\begin{table*}
\centering
\caption{Outlining the parameters defining each galaxy in the simulation catalogue. B/T is the bulge-to-total mass ratio; $b$ is the bulge scale height in kpc; $n$ is the Sersic index; $\lambda'$ describes the stellar component Bullock spin parameter evaluated at $r_{vir}$; $\lambda'$(R$_{\text{eff}}$) is the stellar Bullock parameter evaluated at R$_{\text{eff}}$; $\lambda'_{R}$(R$_{\text{eff}}$) describes the same value corrected using \citeauthor{2007_emsellem}'s conversion, $\lambda' \sim \sqrt{2}/3 \, \lambda'_R$;  $\lambda_R^{90}$ is the value of $\lambda_R$ observed in perfect seeing conditions (i.e. PSF = 0'') at SAMI resolution for each galaxy oriented at 90$^{\circ}$ (edge-on) projected at z = 0.06.}
\label{tab:catalogue}
\begin{tabular}{lccccccc}
\hline
\multicolumn{1}{c}{} & \textbf{B/T} & \textbf{$b$/kpc} & \textbf{n} & \textbf{$\lambda'$} & \multicolumn{1}{l}{\textbf{$\lambda'$(R$_{\text{eff}}$)}} & \textbf{$\lambda'_{R}$(R$_{\text{eff}}$)} & \multicolumn{1}{c}{\textbf{$\lambda_{R}^{\text{90}}$}} \\ \hline
\textbf{S0} & 0.60 & 2.14 & 2.84 & 0.014 & 0.11 & 0.23 & 0.50\\
\textbf{Sa} & 0.40 & 1.38 & 2.26 & 0.022 & 0.16 & 0.34 & 0.60\\
\textbf{Sb} & 0.25 & 0.90 & 1.64 & 0.027 & 0.21 & 0.45 & 0.64\\
\textbf{Sc} & 0.05 & 0.17 & 0.99 & 0.034 & 0.29 & 0.62 & 0.77\\
\textbf{Sd} & 0.02 & 0.07 & 0.97 & 0.035 & 0.31 & 0.66 & 0.78\\ \hline
\end{tabular}
\end{table*}

\begin{figure}
\centering
\includegraphics[width=\columnwidth]{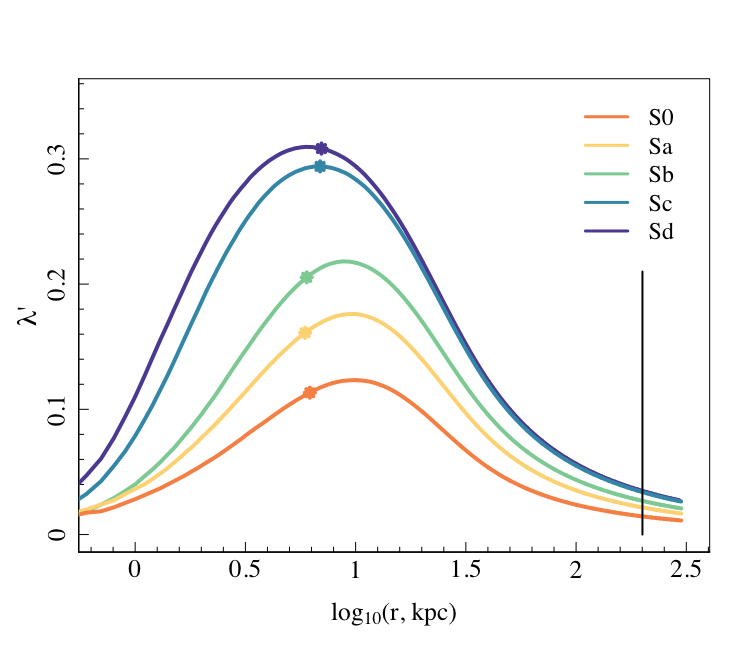}
\caption{Variation of $\lambda'$ of the stellar component with radius for each of the galaxy models. The point at which the Bullock parameter is traditionally measured (the virial radius) is marked at $r_{vir} = 200$ kpc with a black line. The measurements of $\lambda'$ evaluated at the half-mass radius are shown by the star points on each line.}
\label{fig:lambda1}
\end{figure}

For each galaxy in the catalogue, initial conditions are constructed using \galic\, \citep{2014_yurin+springel}. This code takes an iterative approach to solving the collision-less Boltzman equation to create the initial conditions of isolated N-body galaxy models in equilibrium. Stellar bulges are constructed in Hernquist profiles with ergodic velocity structures described by,
\begin{equation}
\rho_{\text{bulge}}(r) = \frac{M_{b}}{2\pi}\frac{b}{r(r + b)^3} \;,
\end{equation}
where $M_{b}$ is the bulge mass, r is the radius and $b$ is the bulge scale length. Stellar disks are formed with an axis-symmetric velocity structure within an exponential profile described by,
\begin{equation}
\rho_{\text{disk}}(R,z) = \frac{M_{d}}{4\pi z_{0} h^2} \text{sech}^2 \left(\frac{z}{z_{0}}\right)\text{exp}\left(-\frac{R}{h}\right) \;,
\end{equation}
where $M_{d}$ is the disk mass, $R$ is the radius in the plane of the disk, $z$ is the height off the plane of the disk, $h$ is the scale length and $z_{0}$ is the scale height. 

To demonstrate that the galaxies are equilibrated, long-lived structures, we evolve them over 10 Gyr using an extended version of the $N$-body/SPH code \gadget\, \citep{2005_springel}, in which the underlying $1.85 \times 10^{12} $M$_{\odot}$ dark matter halo has been replaced with an analytical form of the Hernquist dark matter profile used in \galic. The use of an analytic potential suppresses numerical artefacts (e.g. numerical heating of the disc) that usually arises when there is a mismatch in dark matter and stellar particle mass, as occurs when high resolution is adopted in the disc but not in the halo; further details are given in appendix \ref{sec:appendix_anpot}.

Every galaxy in the catalogue has a stellar mass of $10^{10} $M$_{\odot}$ with the proportion of mass contained in the bulge and disk specified in Table \ref{tab:catalogue}. Each model contains $2.5 \times 10^6$ particles with individual masses of $4 \times 10^3 $M$_{\odot}$ and softening sizes of 100 pc. The stellar disks have radial scale lengths of $h = 4.25$ kpc and scale heights of $z_{0} = 0.85$ kpc. In order to determine the ellipticity of our galaxy models, we compute iso-density contours for each by first generating a flux image at a high resolution (spaxels of 0.05'') projected at 90$^{\circ}$ using \simspin{} with all particles within a sphere of radius $r_{200}$. We use a mass-to-light ratio of 1, such that this flux image is just a map of the mass distribution within the model. The R-package \textsc{ProFound} then takes this image and rank orders the pixels into concentric isophotes containing equal amounts of mass in each. The ellipticity or flattening, $\epsilon = 1 - q$, (where $q$ is the axial ratio of each model) is calculated using the \texttt{ProFound::profoundGetEllipses()} function, which takes all pixels within an isophote containing half the total mass and diagonalizes the inertia tensor to give the axial ratio, $q$ \citep{2018_robotham}. This method ensures that our ellipticities are consistent with what would be measured observationally, as shown in Fig \ref{fig:cat}. The elliptcity, $\epsilon$, is defined at R$_{\text{eff}}$, as in \cite{2007_cappellari, 2016_cappellari}.

We analyse a variety of the inherent kinematic properties of each galaxy to provide comparison to the synthetic observations of $\lambda_R$ later on. $\lambda'$ is evaluated for the stellar component of our models using Eq.~\ref{eq:spin3} and plotted as a function of radius as shown in Fig.~\ref{fig:lambda1}. In \citeauthor{2001_bullock}'s (\citeyear{2001_bullock}) definition, the spin parameter of a dark matter subhalo is calculated at the virial radius of that object. In Fig.~\ref{fig:lambda1}, this radius is shown by the black line. When attempting to make a comparison to observable data, it is very unlikely that the kinematics of a galaxy could be studied out to $\sim 200$ kpc; $\lambda_R$ is typically measured out to one effective radius (on the order of a few kpc). For this reason, we also calculate $\lambda'$ of the stellar component at the half-mass radius, $\lambda'$(R$_{\text{eff}}$). We have assumed that each particle in the simulation has an equal mass-to-light ratio throughout our analysis and hence that the half-mass radius is equivalent to the half-light radius R$_{\rm eff}$; this is the simplest assumption we can make with these basic galaxy models. These half-mass radii are shown in Fig.~\ref{fig:lambda1} by the coloured stars for each galaxy and the associated spin denoted as $\lambda'$(R$_{\text{eff}}$) in Table~\ref{tab:catalogue}. In the derivation of $\lambda_R$ from Eq.~\ref{eq:spin1} in appendix A of \cite{2007_emsellem}, it is shown that $\lambda \sim \sqrt{2}/3 \, \lambda_R$. Conversion factors are required to take account for the differences between one parameter that is defined with full knowledge of the 3D distribution of mass and another that is dependent on a 2D projected light distribution. In appendix \ref{sec:appB}, we show that a similar method can be followed in order to derive $\lambda_R$ from Eq.~\ref{eq:spin2}. We use the same conversion factors to calculate the value we would expect to recover observationally at the effective radius shown as $\lambda'_R$(R$_{\text{eff}}$) for each model in Table~\ref{tab:catalogue}. 

\begin{figure*}
\centering
\includegraphics[width=0.9\textwidth]{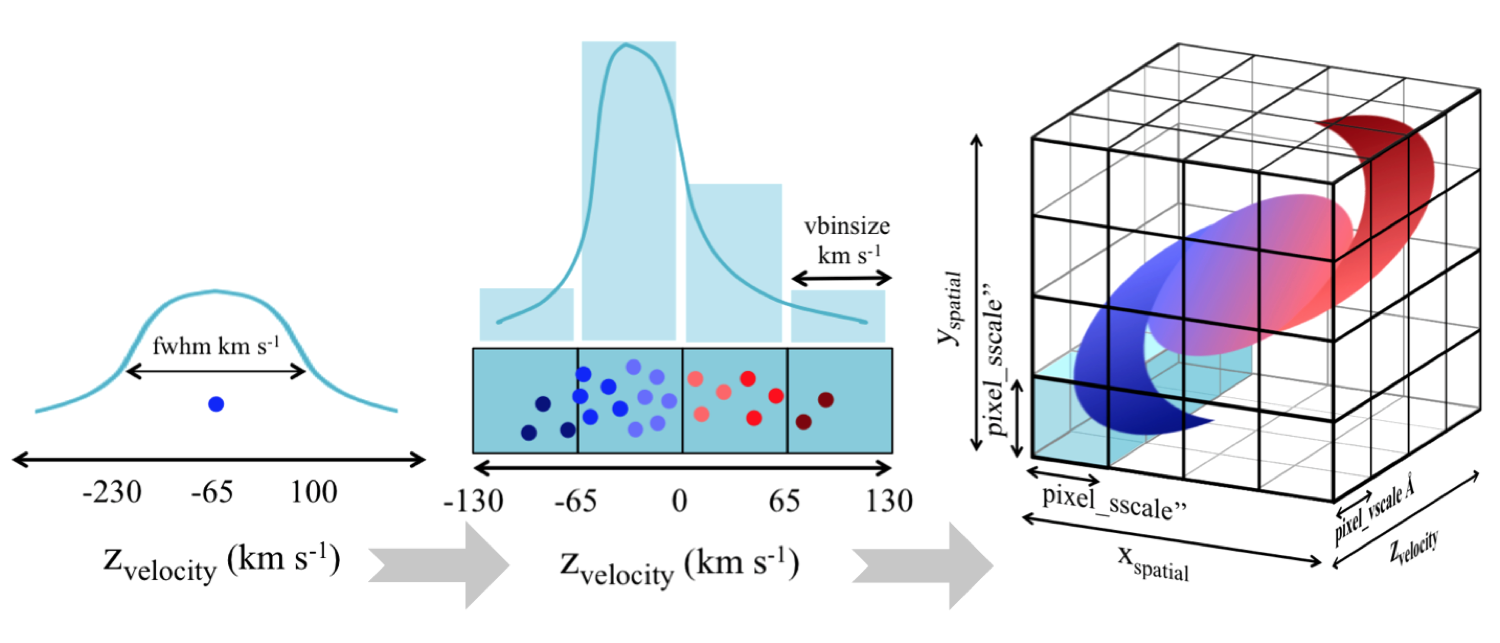}
\caption{The process followed by \simspin\, when constructing a kinematic data cube. Each particle in the simulation has some known velocity, $v_{\text{true}}$. Observationally, we cannot be precise about the velocities we measure. To encode this uncertainty in the mock observation, we assume that each particle's velocity is a Gaussian distribution centred on $v_{\text{true}}$ with a width described by the LSF. If the velocity distribution of each particle extends beyond the width of a single voxel in the cube, fractional contributions will be assigned to each bin accordingly.}
\label{fig:build_data_cube}
\end{figure*}

\begin{figure}
\centering
\includegraphics[width=\columnwidth]{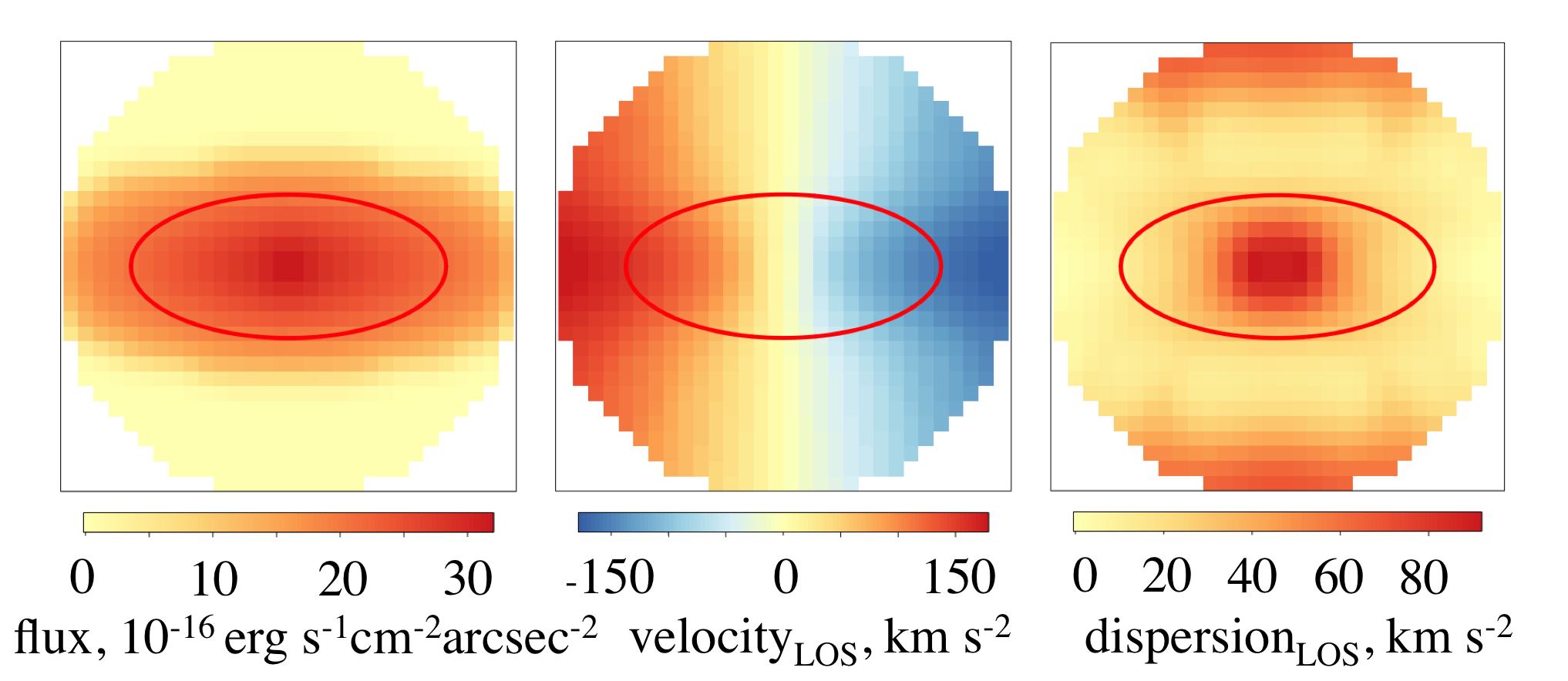}
\caption{Using \simspin, we generate mock IFU images by collapsing the cube in Fig. \ref{fig:build_data_cube} along the z-axis. Here we show the Sd model inclined to 70$^{\circ}$ blurred with a Moffat PSF FWHM = 1'' to mimic seeing effects.}
\label{fig:observation_maps}
\end{figure}

Each of these ``inherent'' spin parameter values may be valid to describe the kinematics of a given galaxy. However, the ambiguity of ``intrinsic'' spin is dangerous. As shown in Table~\ref{tab:catalogue}, every value of the spin parameter belonging to each galaxy is quite different despite the fact that they all provide measures of the specific angular momentum. While, qualitatively, each flavour of spin gives us the same type of information (i.e., disks have a larger spins than the more bulge dominated models), the variation across the Hubble sequence is quantitatively different. When it comes to quantifying scaling relations and linking observations to theory in cosmological simulations, it is important to understand which of these values is actually recovered by observations of $\lambda_R$. It is not computationally efficient to take mock observations of all galaxies in a cosmological volume when kinematic properties like the spin have already been computed. Evidently, from the range of values in Table \ref{tab:catalogue}, it is not sufficient to compare $\lambda'$ and $\lambda_R$ directly without first considering what these values really mean. One of the aims of this work is to examine and present the links between these inherent and observed spin parameters to aid communication between the theoretical and practical communities. This is further discussed in section \ref{sec:def_bias}.

\subsection{The synthetic observations}
\label{sec:observations}

For this experiment, we have generated images based on the parameters of the Sydney-AAO Multi-Object Integral Field Spectrograph (SAMI) which is mounted at the prime focus of the 3.9m Anglo-Australian Telescope. SAMI has 13 imaging fibre bundles, known as ``hexabundles''; within each hexabundle, 61 multimode fibres each subtend a diameter of 1.6'' such that each bundle can view 15'' across its diameter \citep{2011_croom}.  All 793 fibres, plus 26 sky fibres, are fed into the AAOmega dual beam spectrograph comprised of a red and blue arm.  

The majority of the absorption lines used for stellar kinematic measurements fall on the blue arm of the spectrograph and so these parameters are used to generate the mock images for this experiment. For SAMI, a 580V grating is mounted on the blue arm, giving a resolution of R $\sim 1700$ and wavelength coverage of 3700-5700\AA. It has been shown that the line spread function (LSF) describing the spectral instrument response of the prism and gratings on the extracted spectra from the blue arm is well approximated by a Gaussian profile with full-width half-maximum, FWHM$_{\text{blue}}$ = 2.65 \AA $\:$ \citep{2017a_sande}. The kinematic data cubes produced on this arm have a spatial pixel scale of 0.5" and velocity pixel scale of 1.04 \AA $\:$\citep{2017_green}. The flux in each spaxel is mimicked by assigning a luminosity to each particle type using a mass-to-light ratio, $1 \Upsilon_{\odot}$, and approximating the flux from the projection distance.  

Following the evolution of each galaxy model, a series of mock observations were made using the R-package \simspin. Given a \gadget\, snapshot file, a kinematic data cube is constructed using this code according to the SAMI specifications. The spatial and velocity bin (spaxel and voxel) sizes supplied dictate the grid of the cube. Details such as the velocity uncertainty in an observation is incorporated into the simulation by assuming that each particle has a Gaussian distribution of possible velocities with width dictated by the LSF and summed such that portions of each particle's velocity distribution may fall across several voxels. This method is illustrated in Fig. \ref{fig:build_data_cube}. 

The code will go on to generate a series of mock observational images from the kinematic cube. The observed flux, LOS velocity and LOS velocity dispersion is mapped by collapsing the cube down each spaxel. Flux maps are constructed by summing the flux contribution along each spaxel across the various velocities; LOS velocity images are constructed by taking weighted averages of the voxel bins along each spaxel; the LOS dispersion maps are weighted standard deviations of the same voxel distributions along each spaxel. One such example of the images output using this method is shown in Fig. \ref{fig:observation_maps}.  

$\lambda_R$ is then calculated via,
\begin{equation}
\lambda_R = \frac{\sum_{i=1}^{n_b} F_i R_i |V_i|}{\sum_{i=1}^{n_b} F_i R_i \sqrt{V_i^2 + \sigma_i^2}} \;,
\label{eq:obs_lr}
\end{equation}
\noindent where $F_i$ is the observed flux, $R_i$ is the circularised radial position from the centre, $V_i$ is the LOS velocity and $\sigma_i$ is the LOS velocity dispersion per bin, $i$, and summed across the total number of bins, $n_b$, within measurement radius, R$_{\rm eff}$ \citep{2011_emsellem}. R$_{\rm eff}$ is calculated by fitting isophotal ellipses to a flux image generated at the same projected angle but a higher resolution than the IFU image as described in section \ref{sec:setup} - under the assumption that an observer would determine such parameters from a corresponding optical image.

\simspin\, also allows us to investigate the effects of observational limitations on the measurement of $\lambda_R$:  
\begin{enumerate}
\item \emph{Inclination effects} - Projection and distance effects have been measured simply by observing each model across a range of inclinations from $0^\circ$ (face-on) to $90^\circ$ (edge-on) at 5$^\circ$ increments and at a series of redshift distances within the SAMI range 0.04 $<$ z $<$ 0.1. These redshifts were sampled at $\Delta z$ = 0.01 increments. A series of 665 kinematic cubes were generated in total.
\item \emph{Measurement radius effects} - We calculate $\lambda_R$ within a range of measurement radii. From Fig. \ref{fig:lambda1}, it is clear that the spin parameter will change with radius quite significantly for late type and lenticular galaxies. We would like to see how the uncertainty in $\lambda_R$ propagates due to differences in the measurement radius that is used. Over time, the favoured method of determining the effective radius of a galaxy has changed from circular to elliptical apertures \citep{2012_kelvin}, and even more recently via methods of iterative dilation \citep{2018_robotham}. Differences across these methods result in variations in the quoted  R$_{\rm eff}$. In this work, we have considered a range of different radii, from 0.5 - 1.5 R$_{\text{eff}}$, at increments of 0.1 R$_{\rm eff}$. This gives us a further 950 observations to work with across 19 inclinations and 10 additional radii. 
\item \emph{Spatial blurring effects} - ``Beam smearing'' is an inherent problem in Integral Field Units (IFUs) which results in a distortion of the inner rotational velocity curve that is key to measuring $\lambda_R$ \citep{2016_cecil}. Combined with other observational effects, such as atmospheric seeing, we tend to see artificially high levels of LOS velocity dispersion for the central regions of galaxies. We expect this to reduce the measured value of $\lambda_R$, as dispersion appears to be more dominant. These spatial blurring conditions have been modelled in the mock observations by convolving each spatial plane of the data cube with a Moffat point spread function (PSF) at a selection of FWHM values between 0 - 4'', seperated by 0.5'' increments. These observations are again made across a range of inclinations from $0^\circ$ to $90^\circ$ resulting in a further 760 kinematic cubes.
\end{enumerate}

Each of these observation parameters have been tested individually to quantify their impact on the measurement of $\lambda_R$. We can then examine how well the \cite{2018_graham} empirical correction works for an entirely independent set of simulations. This is the secondary aim for the paper: to understand how well we can reverse the effects of observational limitations. 

\section{Results}
\label{sec:results} 

\begin{figure}
\centering
\includegraphics[width=\columnwidth]{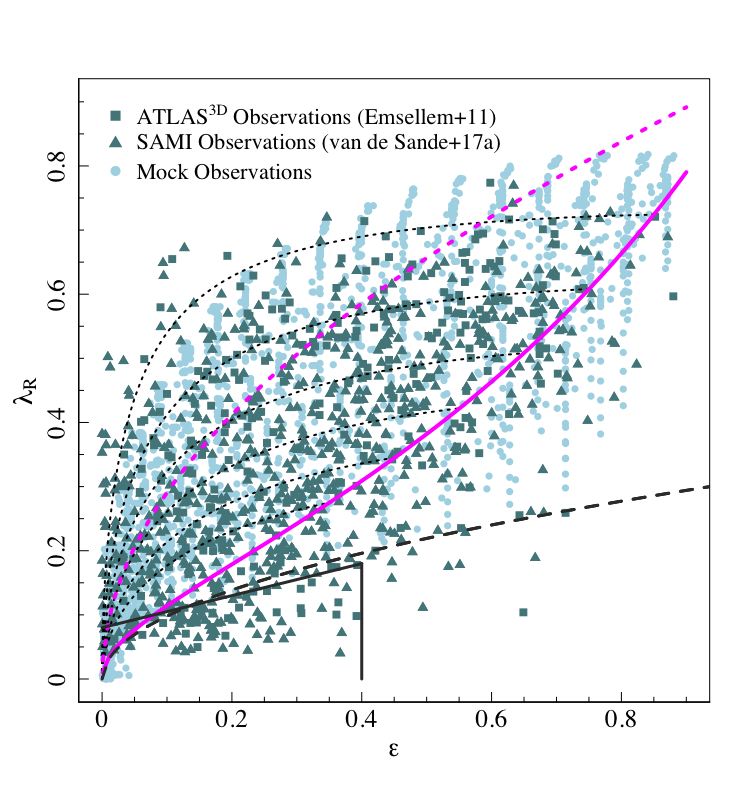}
\caption{Parameter space ($\lambda_R$-$\epsilon$) of all mock observations made in this experiment, as shown in light blue, in comparison to SAMI \citep{2017b_sande} and ATLAS$^{\text{3D}}$ data \citep{2011_emsellem} in dark green. The grey lines show the SR/FR boundaries suggested by \citeauthor{2011_emsellem} in \citeyear{2011_emsellem} (long dashed line), and \citeauthor{2016_cappellari} in \citeyear{2016_cappellari} (solid line). The magenta line shows the edge-on view for oblate galaxies with anisotropy described by, $\beta = 0.7 \times \epsilon$ while the magenta dashed line shows the same relationship for $\beta = 0$ (see \citealt{2007_cappellari}), and black dotted lines show the inclination dependence of galaxies with intrinsic ellipticities, $\epsilon_{intr} = 0.35, 0.45, 0.55, 0.65, 0.75, 0.85$. See text for further details.} 
\label{fig:observations}
\end{figure}

Using \simspin, we have made 2375 mock observations of the 5 galaxy models in total. The results of these investigations are summarised in Fig. \ref{fig:observations}. The parameter space occupied by the synthetic data points is shown in light blue circles. Real observations from ATLAS$^{\text{3D}}$ \citep{2011_emsellem} and SAMI \citep{2017b_sande} are plotted in dark green for comparison. The grey boundary lines distinguish between the galaxies that are classed as slow and fast rotators \citep{2007_emsellem, 2011_emsellem, 2016_cappellari} \footnote{\cite{2007_emsellem} defined a SR such that $\lambda_R < 0.1$, but in \citeyear{2011_emsellem} updated this definition to take into account the larger ATLAS$^{\text{3D}}$ data set such that a SR satisfies $\lambda_R < 0.31 \times \sqrt{\epsilon}$, where $\epsilon$ is the ellipticity of the observed system. More recently, \citet{2016_cappellari} argued that non-regular SRs are better categorised by $\lambda_R < 0.08 + \epsilon_0 / 4$, where $\epsilon_0 < 0.4$, thereby reducing the risk of missing FRs that are very round.}. The magenta lines are a prediction from \cite{2005_binney} for the trend you would expect to see for oblate galaxies of different intrinsic ellipticity ($0 < \epsilon_{intr} < 0.95$) and anisotropy ($\beta$) when measured edge-on. Having calculated the edge-on values of $\lambda_R$, we can then predict the values we would expect at a range of projected inclinations ($0-90^{\circ}$) using,
\begin{equation}
\lambda_R(i) = C(i) \; \frac{\lambda_{R}^{90}}{\sqrt{1 + {\lambda_{R}^{90}}^2(C^2(i) -1)}} \;,
\label{eq:dotted}
\end{equation}
where $C(i) = \text{sin}(i) / \sqrt{1 - \beta \text{cos}^2(i)}$ and $\lambda_{R}^{90}$ is the edge-on value of the observed spin parameter \citep{2007_cappellari, 2007_emsellem}. Here we show the two common extremes of anisotropy, $\beta = 0$ (dashed magenta line) and $= 0.7 \epsilon_{intr}$ (solid magenta line). Using Eq. \ref{eq:dotted}, we plot the black dotted lines in Fig. \ref{fig:observations} for galaxies with intrinsic ellipticities $\epsilon_{intr} = 0.35, 0.45, 0.55, 0.65, 0.75, 0.85$. 

\begin{figure}
\centering
\includegraphics[width=\columnwidth]{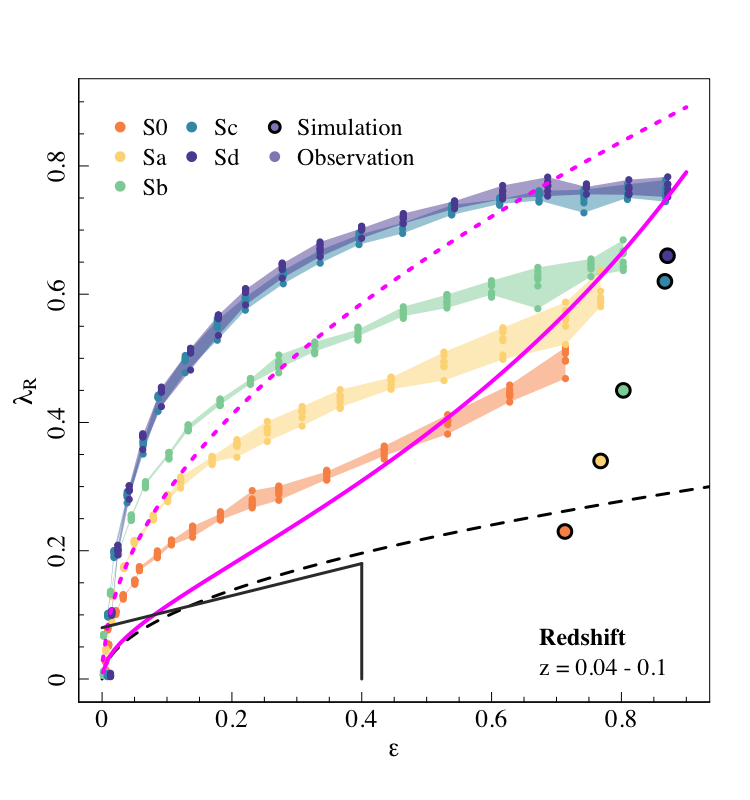}
\caption{The $\lambda_R$ measurement for each galaxy at a range of projected distances from z = 0.04 - 0.1. Each point represents a single mock observation of the model. The spin parameter inherent to the simulation, $\lambda'_{R}$(R$_{\text{eff}}$), as in Table \ref{tab:catalogue}, is shown as the black bordered simulation points. The grey SR/FR boundaries and magenta lines are the same as shown in Fig. \ref{fig:observations}.}
\label{fig:z_e}
\end{figure}

\begin{figure}
\centering
\includegraphics[width=\columnwidth]{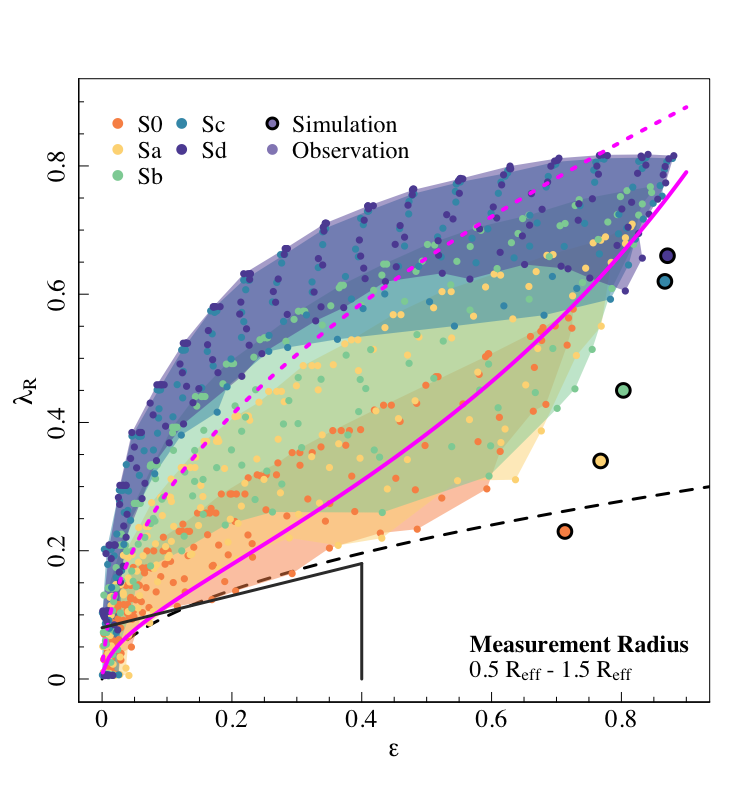}
\caption{Demonstrating the spread in the $\epsilon$-$\lambda_R$ relationship that is caused by varying the measurement radius. Each galaxy has been observed at z = 0.06, through inclinations of $0-90^{\circ}$ and $\lambda_R$ measured within a range of radii. $\lambda'_{R}$(R$_{\text{eff}}$) values calculated per simulation are shown as black bordered points. The grey SR/FR boundaries and magenta lines are the same as shown in Fig. \ref{fig:observations}.}
\label{fig:reff_e}
\end{figure}

\begin{figure*}
\centering
\includegraphics[width=\textwidth]{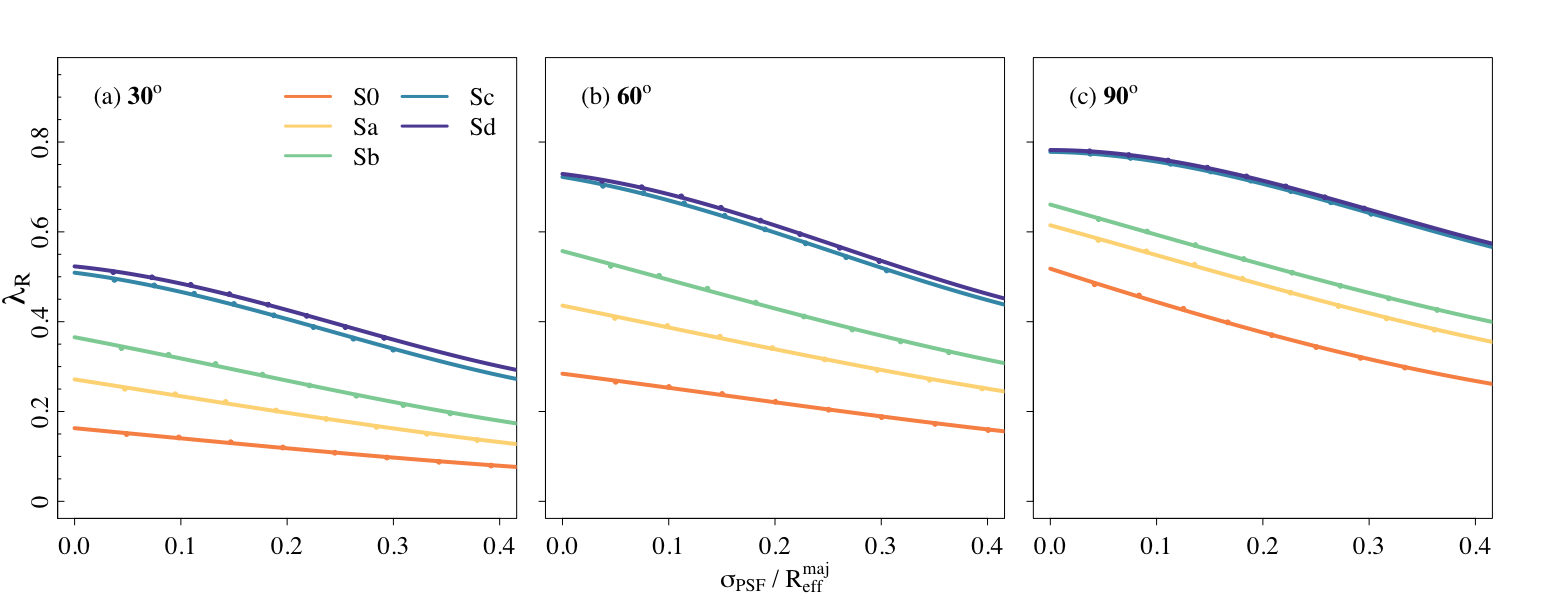}
\caption{The relationship between beam smearing and $\lambda_R$ shown for 3 different projected inclinations, (a) $30^\circ$, (b) $60^\circ$ and (c) $90^\circ$. Each galaxy is observed at a redshift distance of z = 0.06 and $\lambda_R$ measured within 1 R$_{\rm eff}$.}
\label{fig:fwhm}
\end{figure*}

\begin{figure}
\centering
\includegraphics[width=\columnwidth]{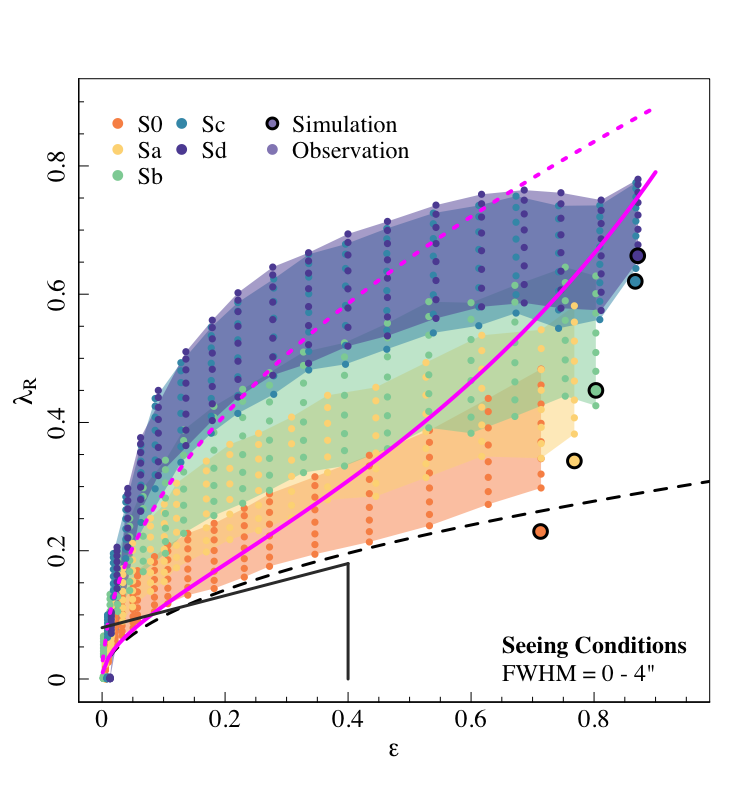}
\caption{Demonstrating the spread in the $\epsilon$-$\lambda_R$ relationship that is caused by varying seeing conditions. Each galaxy has been observed at z = 0.06, through inclinations of $0-90^{\circ}$ and $\lambda_R$ measured within 1 R$_{\rm eff}$. $\lambda'_{R}$(R$_{\text{eff}}$) values calculated per simulation are shown as black bordered points. The grey SR/FR boundaries and magenta lines are the same as shown in Fig. \ref{fig:observations}.}
\label{fig:fwhm_e}
\end{figure}

The area occupied by the true and mock observations overlap very well, specifically in the FR regime. A few galaxies do fall outside our probed region, but this is to be expected. The N-body models that have been measured throughout this work are very simplistic. We have created a series of isolated ergodic bulges and axis-symmetric disks in equilibrium with no gas component. The small number of inaccessible observational examples that sit to the left of our measurements are very round, fast rotators with spiral arms and bars - features which do not appear in this model catalogue due to the specified mass and velocity conditions. The other region we do not sample well is the lower left of the SR regime. Observationally, this region tends to be occupied by non-regular rotators and galaxies with kinematically decoupled cores, which we do not explore in this work. 

Another obvious difference between the measurement of $\lambda_R$ in this work and \citeauthor{2017a_sande}'s work is the definition of radius used in Eq.~\ref{eq:obs_lr}. For the SAMI galaxies in this comparison, the measurement of radius, $R_i$, is defined as the ``semi-major axis of the ellipse on which spaxel $i$ lies'', as opposed to the circularised radius used by \citeauthor{2007_emsellem} in his original work in \citeyear{2007_emsellem} and \citeyear{2011_emsellem} for the ATLAS$^{\text{3D}}$ observations. The purpose of using this elliptical definition of the radius is to remove the inclination dependence of $\lambda_R$. \cite{2017a_sande} show that, for rounder galaxies where $\epsilon < 0.4$, there is very little difference between the spin measured using each radii measure; it becomes more important for the flattened galaxies, though the median difference between circularly measured and intrinsic radii is $\Delta \lambda_R \sim$ 0.04. As we are interested in exploring the effects of projection on the value that is recovered when paired with other limitations, we have chosen to stick with the circularised radius but draw attention to the fact that the definition of $R_{i}$ in Eq.~\ref{eq:obs_lr} may vary across different surveys.

\subsection{Projection effects}
\label{sec:control}

To begin, we made mock observations of each galaxy model across a range of inclinations and projected distances. At this stage, no further observational limitations were included in order to provide a benchmark control by which to compare and to ensure that our analysis performs as expected. This corresponds to a subset of 665 observations of the 5 galaxies. In each case, we use equation \ref{eq:obs_lr} to calculate $\lambda_R$ within an effective radius. 

We found the results of this experiment on the disk galaxies quite surprising. The effect of moving a galaxy further and further away from the observer is that each spaxel in the IFU cube covers a larger portion of the galaxy. This reduced resolution was thought to have a similar effect to increasing the level of spatial blurring; hence, it was expected that $\lambda_R$ would decrease with distance. Instead, we actually see a very slight positive correlation on average across all inclinations between the observed spin parameter and redshift distance.  Increasing the distance from the observer, and hence the spaxel size relative to the galaxy, does not blur the central velocity dispersion in the same way that spatial smearing does. The growth of the spatial pixel means that slightly more disk flux contributes to the edge pixels within $\rm R_{eff}^{maj}$. For the disk galaxies, we find that between z = 0.04 and 0.1, up to 1\% more disk particles fall within the effective radius of the galaxy. This means that additional rotational components are contributing to the measurement, causing the measurement of $\lambda_R$ to rise a very small amount, of order $\sim 0.01$ per $\rm R_{fov}/ R_{eff}^{maj}$. 

Fig. \ref{fig:z_e} shows the results of this test within the observer plane, $\lambda_R$ vs. ellipticity, $\epsilon$. The observed points follow the shape we would expect with inclination \citep{2009_jesseit, 2011_emsellem}. The effects of distance are evident in the models that contain disk components, as shown by the scatter of mock observations.  The level of this effect is negligible - small enough to fall within other observational errors - and so will not effect the measurement of $\lambda_R$. However, it is interesting to note that this is not the negative correlation one might expect.

\subsection{Measurement Radius}
\label{sec:radius}

We briefly consider the impact of measurement radius and its impact on the $\lambda_R$-$\epsilon$ parameter space in Fig. \ref{fig:reff_e}. Here we varied the radius within which $\lambda_R$ was measured at regular increments of $\Delta$R$_{\rm eff}$ = 0.1 from 0.5 - 1.5 R$_{\rm eff}$. All other variables remain fixed. The ellipticity at each radius was calculated by specifying that more or less flux be contained within a given isophote. All pixels within that isophote were then used to find the ellipticity as described in section \ref{sec:setup}. We projected each galaxy from the catalogue through the full range of inclinations, but kept the redshift distance set at z = 0.06 giving a sample of 1045 mock observations.  

Fig. \ref{fig:reff_e} shows that the overlap between the disk models can be significant, and arises because an underestimate of a galaxy's $R_{\rm eff}$ causes its $\lambda_R$ to overlap with the $\lambda_R$ from an overestimated R$_{\rm eff}$ of a galaxy with a slightly larger bulge. This is consistent with recent work \citep[e.g.][]{2017b_sande}.

\subsection{Spatial blurring}
\label{sec:seeing}

The final step was to mimic the effects of beam smearing and atmospheric seeing conditions by introducing spatial blurring to the mock IFU cubes. Each galaxy was projected through the various inclinations, but fixed at a redshift distance of z = 0.06. Following the production of the kinematic IFU data cube, each spatial plane is convolved with a Moffat PSF with FWHM from 0 - 4" at increments of 0.5". Flux, LOS velocity and LOS velocity dispersion images are then constructed as described in section \ref{sec:observations}. $\lambda_R$ is measured from these images within 1 $\rm R_{eff}$ giving a further 760 observations to examine. Note that the effective radius of each model is measured from the unblurred image as we assume that, for real observations, this would not be taken from the IFU data but from higher resolution optical counterparts. 

Applying a Moffat convolution kernel blurs the flux image and smooths out velocity gradients, similar to the effects of spatial blurring seen in real astronomical data. This has the expected result of increasing regions of artificial dispersion causing the recovered value of $\lambda_R$ to fall as seeing conditions grow worse. Across all inclinations, we see that there is a negative correlation between the size of the PSF and $\lambda_R$, as plotted in Fig. \ref{fig:fwhm}. Here we have quantified the level of blurring by the fraction $\rm \sigma_{PSF} / R_{eff}^{maj}$, where $\sigma_{\rm PSF}$\footnote{$\sigma =$ FWHM $/ \; 2 \sqrt{2 \rm ln 2}$} is the standard deviation of the blurring Moffat kernel and $\rm R_{eff}^{maj}$ is the semi-major component of the galaxy's effective radius. The level of blurring with respect to the radius within which the spin parameter is measured is important to readily make comparisons of galaxies observed using different telescopes across different nights with varied seeing conditions.

This negative correlation is inclination and morphology dependent. As the galaxy approaches edge on projections (90$^{\circ}$), the gradient of this relationship is the most severe, with a more linear drop off for galaxies with larger disks. At seeing conditions typical of SAMI observations ($\sim0.3-0.4$ $\rm \sigma_{PSF} / R_{eff}^{maj}$), $\lambda_R$ decreases by as much as $\sim0.05-0.2$ across the range of galaxy types in comparison to the value measured if there was no spatial blurring. This is consistent with the uncertainty quoted by \cite{2017a_sande} in which it was proposed from repeat SAMI observations that adverse seeing conditions could lower measurements of $\lambda_R$ by $0.05-0.1$; our study shows that this bias could be slightly larger for very disky galaxies. Similarly, we agree with the work of \cite{2018_greene} in which two corrections were suggested based on the spin parameter measured. Fig. \ref{fig:fwhm} demonstrates that the impact of seeing is not uniform with galaxy type. \citeauthor{2018_greene} suggest two corrections in which slower galaxies are adjusted by $0.075$ and faster are adjusted by $0.125$. Again, these values fall within our range, but we suggest that the bias could be larger for faster rotators. When compared to similar plots made for MaNGA's JAM models in Appendix C of \cite{2018_graham}, Fig \ref{fig:fwhm} has a very similar distribution. We find that their proposed correction for $\lambda_R$ does a reasonable job of reversing this negative correlation across the various galaxy types, as discussed in section \ref{sec:obs_bias}. 

In Fig. \ref{fig:fwhm_e}, we see the effect of beam smearing and seeing on an observer plane where $\lambda_R$ is plotted against the ellipticity, $\epsilon$. This has the greatest impact on distinguishing the galaxies with large bulges (S0/Sa/Sb). We see that an Sa galaxy with $\gtrsim$ 2'' seeing can become confused with an S0 galaxy observed in near perfect conditions; the same can be said for the Sa/Sb comparison. While our models are simplistic and do not describe the whole range of possible galaxy morphologies, we can still conclude from this overlap that, when left uncorrected, atmospheric seeing can make interpreting a specific morphology from the kinematics alone difficult. Distinguishing between the fastest and slowest FR is obviously a simple task as there is no overlap between the S0 and Sd examples within this wide range of seeing conditions. However, this effect is an important one to bear in mind when making specific conclusions about galaxy comparisons based on this kinematic tracer alone and highlights the need for corrections such as those presented in \cite{2018_graham}.

\section{Discussion}
\label{sec:discuss}
So far, we have considered a range of factors that contribute to uncertainties in measurements of $\lambda_R$. We now investigate the extent to which these uncertainties can be modelled and removed from measurements, focusing on the empirical correction recently proposed by \citeauthor{2018_graham} (\citeyear{2018_graham}). In section \ref{sec:obs_bias}, we investigate how effective this correction is for our models - $N$-body realisations, distinct from the JAM-derived models used in \citet{2018_graham} - with the goal of suggesting an effective way to reduce observational limitations on $\lambda_R$ measurements.

We also examine how the theoretical spin parameters, as measured in a numerical simulation, should be compared most appropriately with the observed spin parameter, $\lambda_R$. When modelling a galaxy's $\lambda_R$, the assumption is that the ``true'' $\lambda_R$ is the one that is recovered in the limit of perfect seeing conditions, but it is vitally important to understand how this ``true'' $\lambda_R$ maps to the ``intrinsic'' theoretical spin parameter, of the kind measured in a numerical simulation. In section \ref{sec:def_bias}, we examine the difference between various definitions of the ``intrinsic'' spin parameter and its observational counterpart.

\subsection{Observational bias}
\label{sec:obs_bias}

\begin{figure*}
\centering
\includegraphics[width=0.90\textwidth]{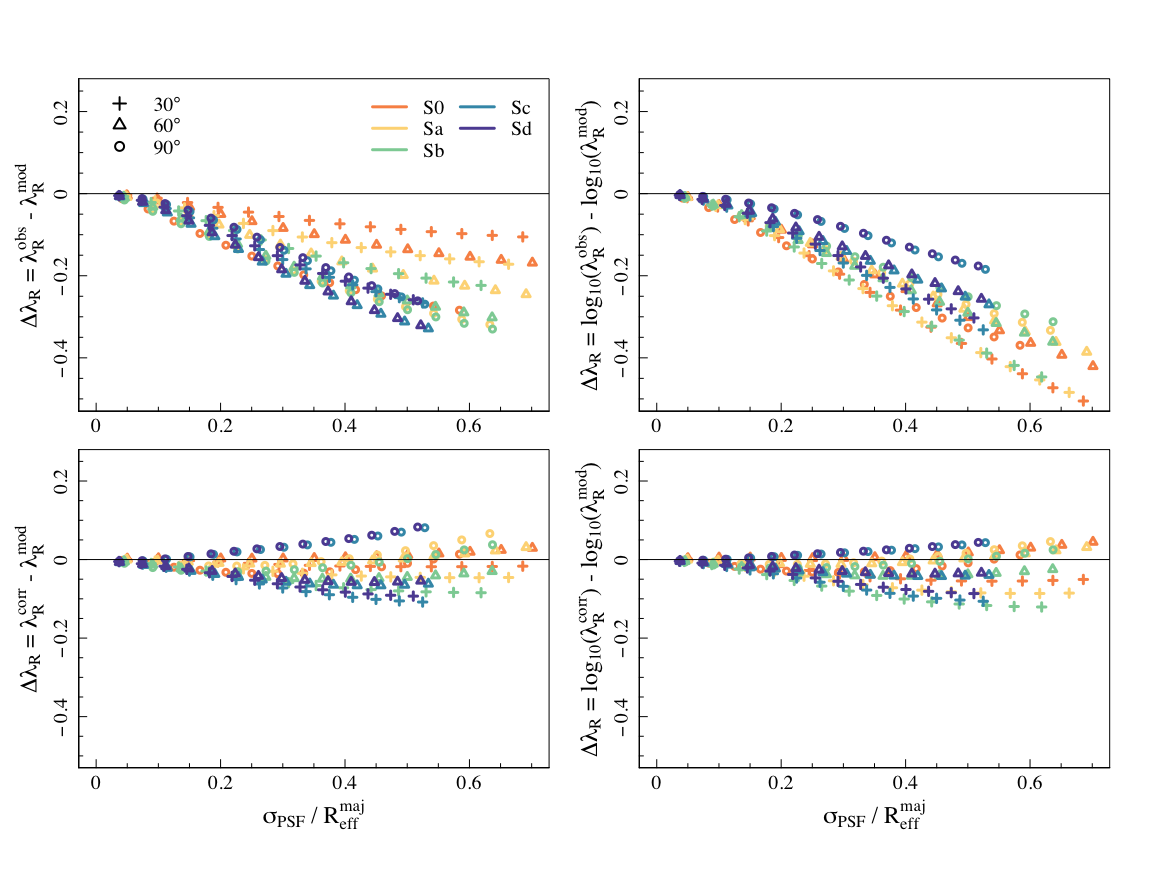}
\caption{Comparing the absolute (left panels) and relative (right panels) differences between the modelled observationally perfect $\lambda_R$ measurements (i.e. no blurring due to seeing), $\lambda_R^{\text{mod}}$, and those observed, $\lambda_R^{\text{obs}}$. The upper panels show $\lambda_R^{\text{obs}} - \lambda_R^{\text{mod}}$, demonstrating the overall effect of increasing the blurring on the observed measurement; the lower panels show  $\lambda_R^{\text{corr}} - \lambda_R^{\text{mod}}$, demonstrating the effect when the observed values are corrected using Eq.~\ref{eq:correction}.}
\label{fig:graham_corr}
\end{figure*}

Recently \cite{2018_graham} presented an empirical formula for correcting $\lambda_R$ to account for the effects of seeing with respect to the semi-major axis of the effective radius. This was derived from a series of 1080 simulations created using the JAM method \citep{2008_cappellari}. We have a further 1425 observations of simulated galaxy models that have been generated using an entirely different method, and so it is interesting to test how effective this formula is for an independent data set. Their proposed equation contains a term to account for the width of the PSF and a term to account for the differences in galaxy morphologies: 
\begin{equation}
\lambda_{R}^{\text{obs}} = \lambda_{R}^{\text{corr}}  g M_2 \left(\frac{\sigma_{\text{PSF}}}{R_{\text{eff}}^{\text{maj}}}\right) f_n \left( \frac{\sigma_{\text{PSF}}}{R_{\text{eff}}^{\text{maj}}} \right) \;,
\label{eq:correction}
\end{equation}
where,
\begin{equation}
g M_2 \left(\frac{\sigma_{\text{PSF}}}{R_{\text{eff}}^{\text{maj}}}\right) = \left[1 + \left(\frac{\sigma_{\text{PSF}}/R_{\text{eff}}^{\text{maj}}}{0.47}\right)^{1.76} \right]^{-0.84} \;,
\end{equation}
is a generalised form of the Moffat function and
\begin{equation}
f_n \left( \frac{\sigma_{\text{PSF}}}{R_{\text{eff}}^{\text{maj}}} \right) = \left[1 + (n-2)\left(0.26\frac{\sigma_{\text{PSF}}}{R_{\text{eff}}^{\text{maj}}}\right)\right]^{-1} \;,
\end{equation}
is an empirical relationship to account for morphological type via observed Sersic indices. 

First, we measure the Sersic indices of our models using the Bayesian galaxy fitting tool, ProFit \citep{2017_robotham}. Isophotal ellipses are extracted from a high resolution flux image of each galaxy model and a single component Sersic profile is fitted to the surface brightness profile produced. These values are shown in Table~\ref{tab:catalogue} and span a reasonable range from $0.9 \lesssim n \lesssim 3$. Following the same method as section \ref{sec:seeing}, we have projected our galaxies at z = 0.06 but increased the level of blurring from 0 - 7'' in this case such that we can investigate a similar range in $\sigma_{\text{PSF}}/$R$_{\text{eff}}^{\text{maj}}$ as \cite{2018_graham}. The correction has then been applied to our synthetic observations using Eq.~\ref{eq:correction} to solve for $\lambda_R^{\text{corr}}$. 

In Fig.~\ref{fig:graham_corr}, we present the results of this correction on our data. In the left lower panel we plot the absolute difference between the modelled, observationally perfect value of $\lambda_R$ ($\lambda_R^{\text{mod}}$) and the value observed following correction ($\lambda_R^{\text{corr}}$) across a range of seeing conditions described by the fraction $\rm \sigma_{PSF} / R_{eff}^{maj}$. We show the absolute difference between $\lambda_R^{\text{mod}}$ and the un-corrected observed value ($\lambda_R^{\text{obs}}$) in the left upper panel as reference to demonstrate the improvement achieved by using this correction. This is shown for each galaxy morphology inclined to $30^{\circ}$, $60^{\circ}$ and $90^{\circ}$. Prior to correction, we can see that the observed value is always an under-estimate, though the amount by which this value is reduced is dependent on the galaxy morphology (with a minor dependence on the observed inclination). The absolute difference between the modelled and observed value generally increases with disk dominance at each seeing increment. As demonstrated by the spread of the points about the zero line in the lower panel, we see a reduction in the absolute uncertainty across all models following correction. The systematic differences between morphological types have been mostly removed. The error on the corrected values is greatly reduced from $\Delta \lambda_R \lesssim 0.2$ to $\lesssim 0.07$ at the typical SAMI seeing of $\sim 0.3-0.4 \, \rm \sigma_{PSF} / R_{eff}^{maj}$. 

If instead we consider the relative differences as shown on the right, using the log difference log$_{10}(\lambda_R^{\text{obs}}) - $log$_{10}(\lambda_R^{\text{mod}})$, we find that this relationship between uncertainty and disk dominance is inverted. While the absolute difference between the measured and inherent properties may be largest for the diskiest galaxies, relatively the uncertainty associated with a more bulge dominated system could be as large as the inherent value itself. Hence, it is very important to consider applying these kinds of corrections across both slow and fast rotator regimes. We see that in the corrected plane, irrespective of whether we consider the absolute or relative uncertainty, this morphological dependence is removed. 

Overall, this shows that the correction presented by \citeauthor{2018_graham} (\citeyear{2018_graham}) is largely successful for an entirely independent set of models. We do see that this is dependent on inclination and agree that this correction will be most effective when applied to galaxies inclined to intermediate angles $i \sim 50$. Given this result, we suggest that this correction is applied to data across all morphological types, especially in comparisons where seeing conditions vary considerably between observations.

\subsection{Definition bias}
\label{sec:def_bias}

\begin{figure*}
\centering
\includegraphics[width=0.90\textwidth]{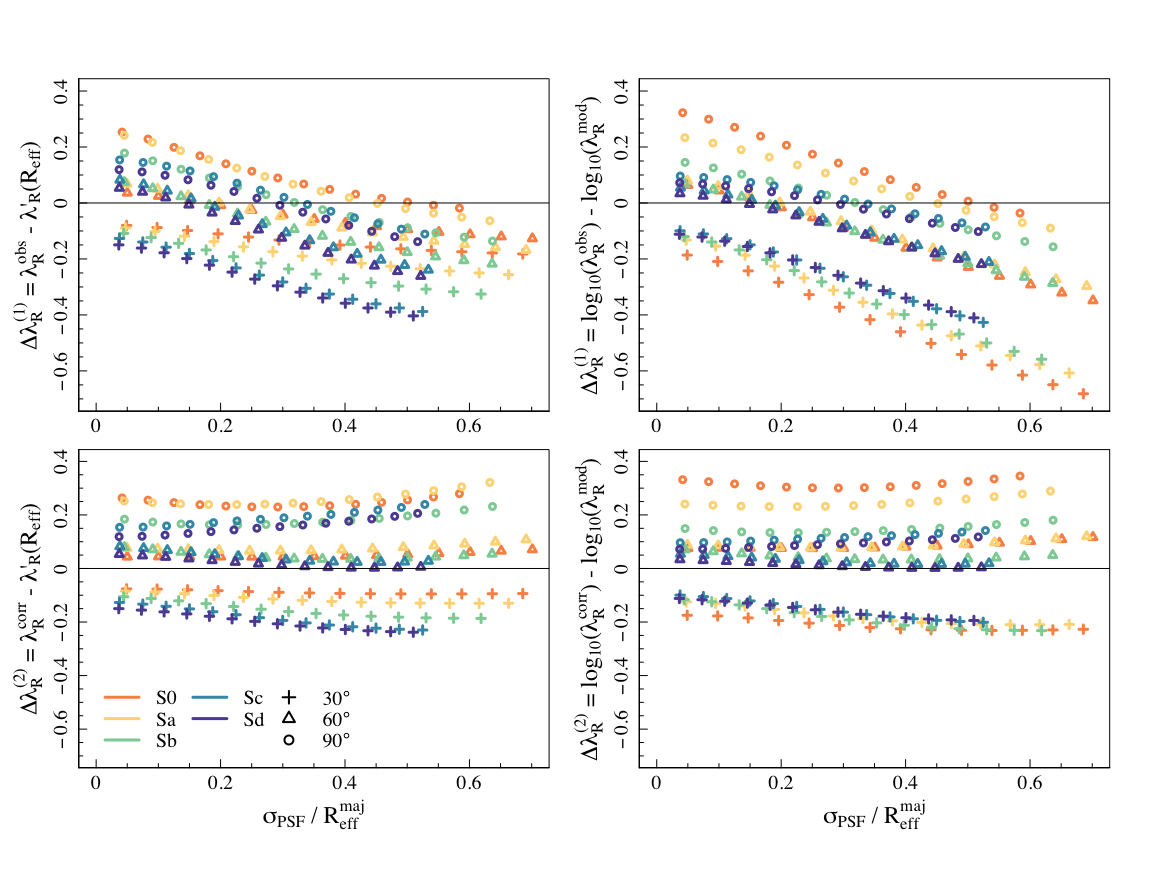}
\caption{Comparing the absolute (left panels) and relative (right panels) differences between the observed and corrected values of $\lambda_R$ with the $\lambda'_R$ value inherent to each simulation at a range of $\sigma_{\text{PSF}} /$ R$_{\text{eff}}^{\text{maj}}$ values for three different projections at $30^{\circ}$ (crosses), $60^{\circ}$ (triangles) and $90^{\circ}$ (circles). On the left, $\Delta \lambda_R^{(1)}$ is the difference between the observed value from the value proposed by \citeauthor{2007_emsellem} (\citeyear{2007_emsellem}) where $\lambda'_R \sim 3/\sqrt{2} \, \lambda'$ while $\Delta \lambda_R^{(2)}$ is the same difference but taken from the observed value that has been corrected for seeing using \citeauthor{2018_graham}'s (\citeyear{2018_graham}) correction.}
\label{fig:delta_lr_PSF}
\end{figure*}

Aside from the effects of the atmosphere on our observations, there is also the issue of definition bias that makes it difficult to compare spin parameters across real and simulated data sets. Often the ``true'' spin parameter is taken to be the one that would be observationally measured if seeing conditions were perfect (for example, \citealt{2018_graham}, \citealt{2018_greene}, \citealt{2017b_sande}, \citealt{2013_deugenio}), while in simulations the Bullock parameter is used \citep{2015_teklu,2017_rodriguez, 2017_zjupa,2017_stewart}. This study has shown that the effects of projection, distance, seeing and measurement radius will impact the recovered value of $\lambda_R$, but even in ideal conditions this value may not be a direct proxy for the spin of a galaxy that a theorist would measure within a cosmological simulation. 

Because we have the 3D model of the galaxy at our disposal, we can explore the inherent property of the system that $\lambda_R$ is actually measuring. We describe the intrinsic property of the system to be the value of the stellar Bullock spin parameter $\lambda'$ that would be recovered from a simulation, as in \cite{2015_teklu}. In order to make the comparison between observation and simulation, we must account for the fact that usually $\lambda'$ is evaluated at $r_{200}$, rather than R$_{\text{eff}}$. In Fig \ref{fig:lambda1} and Table \ref{tab:catalogue}, we demonstrate that the latter values are an order of magnitude larger in general. As shown in appendix \ref{sec:appB}, we also must account for observational projection effects by the scaling $\lambda' \sim \sqrt{2}/3 \lambda_R$.

Fig. \ref{fig:delta_lr_PSF} shows $\Delta \lambda_R$, the absolute (left) and relative differences (right) between the observed $\lambda_R$ value and the stellar Bullock parameter evaluated at R$_{\text{eff}}$ and scaled with \cite{2007_emsellem}'s correction. We show the observed value before (above) and after (below) the correction presented by \cite{2018_graham}. $\Delta \lambda_R$ is shown with respect to the level of blurring, $\sigma_{\text{PSF}}/$R$_{\text{eff}}^{\text{maj}}$. As, before, we have observed each galaxy at a redshift of z = 0.06 up to a PSF of 7'' in order to consider a wider impact of typical seeing conditions. We show the trends for each galaxy model inclined to $30^{\circ}, 60^{\circ}$ and $90^{\circ}$ as crosses, triangles and circles respectively. 

The purpose of this comparison has been to provide a reference between the commonly defined theoretical spin parameter and $\lambda_R$. Having adjusted the observed $\lambda_R$ using the correction provided by \cite{2018_graham}, Fig. \ref{fig:delta_lr_PSF} demonstrates the offsets you may expect between the observed and the stellar Bullock spin parameter. While \citeauthor{2018_graham}'s correction was originally designed to reduce the difference between the observationally perfect value and those effected by seeing, it clearly does a good job at straightening the $\lambda'_R$ relation across all seeing conditions. However, there is still quite a significant uncertainty due to the effects of inclination. Our inherent measure of spin takes into account the 3D distribution of particles in our model, but it is not possible to recover this from the projected, observable $\lambda_R$. Overall, we find that $\lambda_R$ does a fair job of tracking the inherent value proposed by \cite{2007_emsellem}, $\lambda'_{R}(\text{R}_{\text{eff}})$, especially at intermediate inclinations. However, the spread across the $\Delta \lambda_R^{(2)}$ space is as high as $\pm 0.2$.

This analysis makes clear that care must be taken when comparing a \citet{2001_bullock} $\lambda'$ measured from a simulated galaxy and the observationally deduced $\lambda_R$. As our results show, $\lambda_R$ provides a reasonable approximation to $\lambda'$ evaluated for the galactic stellar component within 1 R$_{\rm eff}$ and scaled by \citeauthor{2007_emsellem}'s $3/\sqrt{2}$; this is arguably the most attractive approach because it is straightforward to compute directly from simulation particle data. Perhaps more subtly, Fig \ref{fig:delta_lr_PSF} also demonstrates that at fixed inclinations there are still variations in the absolute and relative $\Delta \lambda_R$ offsets due to morpohology. This indicates that $\lambda_R$ does not map equally to spin for different galaxy types, which has potential implications for the quantification of dynamical scaling relations from IFS data. For example, a slope for the observed mass-angular momentum-spin plane inconsistent with theoretical predictions may not necessarily imply a disagreement between real and simulated galaxies, or viceversa \citep[e.g.,][]{2016_cortese}. Constructing mock IFS observations, deducing $\lambda_R$ from these maps and comparing to observations therefore provides a far more accurate yet more expensive solution.

\section{Conclusions}
\label{sec:conclusions}

The goal of this paper has been to understand the relationship between the observational spin parameter, $\lambda_R$, and theoretical \citet{2001_bullock} spin parameter, $\lambda'$, which is one of the most widely used definitions in numerical simulations. Using $N$-body models of galaxies and mock IFS observations of these data, we have measured $\lambda_R$ as an observer would - incorporating the effects of, for example, beam smearing and seeing - and demonstrated that the empirical correction proposed by \citet{2018_graham} is an effective approach to removing these biases. We have also shown that $\lambda_R$ provides a good approximation to the intrinsic theoretical spin parameter of a galaxy, provided observational biases are corrected for. Measured directly from simulation data, $\lambda'$ evaluated for the galactic stellar component within 1 R$_{\rm eff}$ and scaled by \citeauthor{2007_emsellem}'s $3/\sqrt{2}$ provides a good approximation to the observed $\lambda_R$; mock IFS observations provide greater accuracy, albeit at greater computational cost and complexity of analysis.

We have found, in agreement with several previous studies, that there is a strong negative correlation between $\lambda_R$ and seeing conditions \citep{2013_deugenio, 2017a_sande, 2018_greene, 2018_graham}. We find that at seeing conditions typical of SAMI observations ($\sim0.3-0.4$ $\rm \sigma_{PSF} / R_{eff}^{maj}$), $\lambda_R$ may decreases by as much as $\sim0.05-0.2$, though this reduction is not consistent across all galaxy types. Dispersion dominated systems are affected by seeing to a much lesser degree if we consider the absolute difference between the blurred and un-blurred values. The necessary correction grows with disk dominance.

Finally, we have evaluated the success of \citeauthor{2018_graham}'s (\citeyear{2018_graham}) empirical correction to account for observational bias. Using our independent set of N-body models, we confirm that this formula reduces the effects of seeing from $\Delta \lambda_R \lesssim 0.2$ to $\lesssim 0.07$. There remains a slight inclination dependence however, such that the correction works best for galaxies inclined at intermediate angles, $i \sim 50^{\circ}$. We conclude that this correction should be applied to observational data prior to further comparisons in order to significantly reduce the observational limitations. We have also discussed that, following this correction, the observed $\lambda_R$ may be compared to the theoretical definition of the Bullock spin parameter, $\lambda'$. While often it is assumed that the ``true'' spin parameter is the one measured in perfect seeing conditions, and this empirical correction has been designed to assume as much, we show that $\lambda_R$ is offset from the commonly used theoretical Bullock parameter. We demonstrate that using \citeauthor{2007_emsellem}'s (\citeyear{2007_emsellem}) linear correction to account for projection effects reduces this offset successfully. However, when corrected for seeing, $\lambda_R^{\rm corr}$ is still offset from the inherent $\lambda'_{R}$ specific to the stellar mass measured within a radius R$_{\text{eff}}$ by as much as $\Delta \sim \pm 0.2$ and this offset is not equal for fixed inclination across different morphologies. For the most appropriate comparison without relying on forward modelling techniques, we suggest using the scaled $\lambda'$ value, $\lambda'_{R}$(R$_{\text{eff}}$), with respect to the corrected observations, $\lambda_R^{\rm corr}$ measured at intermediate inclinations.

\section*{Acknowledgements}

The authors thank the anonymous referee for constructive and insightful comments that helped to improve this paper. KH is supported by the SIRF and UPA awarded by the University of Western Australia Scholarships Committee. CP is supported by ARC Future Fellowship FT130100041. AR acknowledges the support of ARC Discovery Project grant DP140100395. DT acknowledges support from a 2016 University of Western Australia Research Collaboration Award. This research was conducted by the ARC Centre of Excellence for All-sky Astrophysics (CAASTRO), through project number CE110001020. Parts of this research were also conducted by the Australian Research Council Centre of Excellence for All Sky Astrophysics in 3 Dimensions (ASTRO 3D), through project number CE170100013. This research was undertaken on Raijin, the NCI National Facility in Canberra, Australia, which is supported by the Australian commonwealth Government and on Magnus at the Pawsey Supercomputing Centre in Perth, Australia. 




\bibliographystyle{mnras}
\bibliography{numerical-twist-spin} 

\begin{thebibliography}{}
\makeatletter
\relax
\def\mn@urlcharsother{\let\do\@makeother \do\$\do\&\do\#\do\^\do\_\do\%\do\~}
\def\mn@doi{\begingroup\mn@urlcharsother \@ifnextchar [ {\mn@doi@}
  {\mn@doi@[]}}
\def\mn@doi@[#1]#2{\def\@tempa{#1}\ifx\@tempa\@empty \href
  {http://dx.doi.org/#2} {doi:#2}\else \href {http://dx.doi.org/#2} {#1}\fi
  \endgroup}
\def\mn@eprint#1#2{\mn@eprint@#1:#2::\@nil}
\def\mn@eprint@arXiv#1{\href {http://arxiv.org/abs/#1} {{\tt arXiv:#1}}}
\def\mn@eprint@dblp#1{\href {http://dblp.uni-trier.de/rec/bibtex/#1.xml}
  {dblp:#1}}
\def\mn@eprint@#1:#2:#3:#4\@nil{\def\@tempa {#1}\def\@tempb {#2}\def\@tempc
  {#3}\ifx \@tempc \@empty \let \@tempc \@tempb \let \@tempb \@tempa \fi \ifx
  \@tempb \@empty \def\@tempb {arXiv}\fi \@ifundefined
  {mn@eprint@\@tempb}{\@tempb:\@tempc}{\expandafter \expandafter \csname
  mn@eprint@\@tempb\endcsname \expandafter{\@tempc}}}

\bibitem[\protect\citeauthoryear{Arnold, Romanowsky, Brodie  et~al.}{Arnold
  et~al.}{2014}]{2014_arnold}
Arnold J.~A.,  Romanowsky A.~J.,  Brodie J.~P.,   et~al., 2014, ApJ, 791, 27pp

\bibitem[\protect\citeauthoryear{Bender}{Bender}{1988}]{1988_bender}
Bender R.,  1988, Astronomy and Astrophysics, 193, L7

\bibitem[\protect\citeauthoryear{Binney}{Binney}{2005}]{2005_binney}
Binney J.,  2005, MNRAS, 363, 937

\bibitem[\protect\citeauthoryear{Brough, van~de Sande, Owers  et~al.}{Brough
  et~al.}{2017}]{2017_brough}
Brough S.,  van~de Sande J.,  Owers M.~S.,   et~al., 2017, ApJ, 844, 12

\bibitem[\protect\citeauthoryear{Bullock, Dekel, Kolatt, Kravtsov, Klypin,
  Porciani  \& Primack}{Bullock et~al.}{2001}]{2001_bullock}
Bullock J.~S.,  Dekel A.,  Kolatt T.~S.,  Kravtsov A.~V.,  Klypin A.~A.,
  Porciani C.,   Primack J.~R.,  2001, ApJ, 555, 240

\bibitem[\protect\citeauthoryear{Bundy, Bershady, Law  et~al.}{Bundy
  et~al.}{2015}]{2015_bundy}
Bundy K.,  Bershady M.~A.,  Law D.~R.,   et~al., 2015, ApJ, 798, 1

\bibitem[\protect\citeauthoryear{Cappellari}{Cappellari}{2008}]{2008_cappellari}
Cappellari M.,  2008, MNRAS, 390, 71

\bibitem[\protect\citeauthoryear{Cappellari}{Cappellari}{2016}]{2016_cappellari}
Cappellari M.,  2016, Annual Review of Astronomy and Astrophysics, 54, 597

\bibitem[\protect\citeauthoryear{Cappellari, Emsellem, Bacon
  et~al.}{Cappellari et~al.}{2007}]{2007_cappellari}
Cappellari M.,  Emsellem E.,  Bacon R.,   et~al., 2007, MNRAS, 379, 418

\bibitem[\protect\citeauthoryear{Cappellari, Emsellem, Krajnovic
  et~al.}{Cappellari et~al.}{2011}]{2011_cappellari_b}
Cappellari M.,  Emsellem E.,  Krajnovic D.,   et~al., 2011, MNRAS, 413, 813

\bibitem[\protect\citeauthoryear{Cecil, Fogarty, Richards  et~al.}{Cecil
  et~al.}{2016}]{2016_cecil}
Cecil G.,  Fogarty L.,  Richards S.,   et~al., 2016, MNRAS, 456, 1299

\bibitem[\protect\citeauthoryear{Conselice}{Conselice}{2014}]{2014_conselice}
Conselice C.~J.,  2014, Annual Review of Astronomy and Astrophysics, 52, 291

\bibitem[\protect\citeauthoryear{Cortese, Fogarty, Bekki  et~al.}{Cortese
  et~al.}{2016}]{2016_cortese}
Cortese L.,  Fogarty L.,  Bekki K.,   et~al., 2016, MNRAS, 463, 170

\bibitem[\protect\citeauthoryear{Croom, Lawrence, Bland-Hawthorn  et~al.}{Croom
  et~al.}{2011}]{2011_croom}
Croom S.~M.,  Lawrence J.~S.,  Bland-Hawthorn J.,   et~al., 2011, MNRAS, 421,
  872

\bibitem[\protect\citeauthoryear{D'Eugenio, Houghton, Davies  et~al.}{D'Eugenio
  et~al.}{2013}]{2013_deugenio}
D'Eugenio F.,  Houghton R. C.~W.,  Davies R.~L.,   et~al., 2013, MNRAS, 429,
  1258

\bibitem[\protect\citeauthoryear{Davies, Efstathiou, Fall  et~al.}{Davies
  et~al.}{1983}]{1983_davies}
Davies R.,  Efstathiou G.,  Fall M.,   et~al., 1983, ApJ, 266, 41

\bibitem[\protect\citeauthoryear{Emsellem, Cappellari, Peletier
  et~al.}{Emsellem et~al.}{2004}]{2004_emsellem}
Emsellem E.,  Cappellari M.,  Peletier R.~F.,   et~al., 2004, MNRAS, 352, 721

\bibitem[\protect\citeauthoryear{Emsellem, Cappellari, Krajnovic
  et~al.}{Emsellem et~al.}{2007}]{2007_emsellem}
Emsellem E.,  Cappellari M.,  Krajnovic D.,   et~al., 2007, MNRAS, 379, 401

\bibitem[\protect\citeauthoryear{Emsellem, Cappellari, Krajnovic
  et~al.}{Emsellem et~al.}{2011}]{2011_emsellem}
Emsellem E.,  Cappellari M.,  Krajnovic D.,   et~al., 2011, MNRAS, 414, 888

\bibitem[\protect\citeauthoryear{Fall \& Efstathiou}{Fall \&
  Efstathiou}{1980}]{1980_fall}
Fall M.~S.,  Efstathiou G.,  1980, MNRAS, 193, 189

\bibitem[\protect\citeauthoryear{Fogarty, Scott, Owers  et~al.}{Fogarty
  et~al.}{2015}]{2015_fogarty}
Fogarty L.,  Scott N.,  Owers M.,   et~al., 2015, MNRAS, 454, 2050

\bibitem[\protect\citeauthoryear{Foster, Arnold, Forbes  et~al.}{Foster
  et~al.}{2013}]{2013_foster}
Foster C.,  Arnold J.,  Forbes D.,   et~al., 2013, MNRAS, 435, 3587

\bibitem[\protect\citeauthoryear{Genel, Fall, Hernquist  et~al.}{Genel
  et~al.}{2015}]{2015_genel}
Genel S.,  Fall M.,  Hernquist L.,   et~al., 2015, ApJ, 804, 7pp

\bibitem[\protect\citeauthoryear{Graham, Cappellari, Li  et~al.}{Graham
  et~al.}{2018}]{2018_graham}
Graham M.~T.,  Cappellari M.,  Li H.,   et~al., 2018, MNRAS, sty504

\bibitem[\protect\citeauthoryear{Green, Croom, Scott  et~al.}{Green
  et~al.}{2017}]{2017_green}
Green A.~W.,  Croom S.~M.,  Scott N.,   et~al., 2017, MNRAS

\bibitem[\protect\citeauthoryear{Greene, Leathaud, Emsellem  et~al.}{Greene
  et~al.}{2018}]{2018_greene}
Greene J.~E.,  Leathaud A.,  Emsellem E.,   et~al., 2018, ApJ, 852, 36

\bibitem[\protect\citeauthoryear{Hubble}{Hubble}{1926}]{1926_hubble}
Hubble E.,  1926, ApJ, 64, 321

\bibitem[\protect\citeauthoryear{Illingworth}{Illingworth}{1977}]{1977_illingworth}
Illingworth G.,  1977, ApJ, 218, L43

\bibitem[\protect\citeauthoryear{Jesseit, Cappellari, Naab  et~al.}{Jesseit
  et~al.}{2009}]{2009_jesseit}
Jesseit R.,  Cappellari M.,  Naab T.,   et~al., 2009, MNRAS, 397, 1202

\bibitem[\protect\citeauthoryear{Kelvin, Driver, Robotham  et~al.}{Kelvin
  et~al.}{2012}]{2012_kelvin}
Kelvin L.~S.,  Driver S.~P.,  Robotham A.~S.,   et~al., 2012, MNRAS, 421, 1007

\bibitem[\protect\citeauthoryear{Knebe \& Power}{Knebe \&
  Power}{2008}]{2008_knebe}
Knebe A.,  Power C.,  2008, ApJ, 678, 621

\bibitem[\protect\citeauthoryear{Lagos, Stevens, Bower  et~al.}{Lagos
  et~al.}{2018}]{2018_lagos}
Lagos C. d.~P.,  Stevens A. R.~H.,  Bower R.~G.,   et~al., 2018, MNRAS, 473,
  4956

\bibitem[\protect\citeauthoryear{Mo, Mao  \& White}{Mo et~al.}{1998}]{1998_mo}
Mo H.~J.,  Mao S.,   White S. D.~M.,  1998, MNRAS, 295, 319

\bibitem[\protect\citeauthoryear{Pedrosa \& Tissera}{Pedrosa \&
  Tissera}{2015}]{2015_pedrossa}
Pedrosa S.~E.,  Tissera P.~B.,  2015, Astronomy and Astrophysics, 584, 8

\bibitem[\protect\citeauthoryear{Peebles}{Peebles}{1969}]{1969_peebles}
Peebles P. J.~E.,  1969, ApJ, 155, 393

\bibitem[\protect\citeauthoryear{{Power}, {Knebe}  \& {Knollmann}}{{Power}
  et~al.}{2012}]{2012_power}
{Power} C.,  {Knebe} A.,   {Knollmann} S.~R.,  2012, \mn@doi [\mnras]
  {10.1111/j.1365-2966.2011.19820.x}, \href
  {http://adsabs.harvard.edu/abs/2012MNRAS.419.1576P} {419, 1576}

\bibitem[\protect\citeauthoryear{Robotham, Taranu, Tobar  et~al.}{Robotham
  et~al.}{2017}]{2017_robotham}
Robotham A.,  Taranu D.,  Tobar R.,   et~al., 2017, MNRAS, 466, 1513

\bibitem[\protect\citeauthoryear{Robotham, Davies, Driver  et~al.}{Robotham
  et~al.}{2018}]{2018_robotham}
Robotham A.,  Davies L.,  Driver S.,   et~al., 2018, MNRAS, 476, 3137

\bibitem[\protect\citeauthoryear{Rodriguez-Gomez, Sales, Genel
  et~al.}{Rodriguez-Gomez et~al.}{2017}]{2017_rodriguez}
Rodriguez-Gomez V.,  Sales L.~V.,  Genel S.,   et~al., 2017, MNRAS, 467, 3083

\bibitem[\protect\citeauthoryear{S{\'a}nchez, Kennicutt, Gil~de Paz
  et~al.}{S{\'a}nchez et~al.}{2012}]{2012_sanchez}
S{\'a}nchez S.~F.,  Kennicutt R.~C.,  Gil~de Paz A.,   et~al., 2012, Astronomy
  and Astrophysics, 538, 33

\bibitem[\protect\citeauthoryear{Sandage, Sandage  \& Kristian}{Sandage
  et~al.}{1975}]{1975_sandage}
Sandage A.,  Sandage M.,   Kristian J.,  1975, Galaxies and the Universe
  (Volume IX of Stars and Stellar Systems).
 Stars and Stellar Systems Vol. 9, University of Chicago Press

\bibitem[\protect\citeauthoryear{{Schaye} et~al.,}{{Schaye}
  et~al.}{2015}]{2015_schaye}
{Schaye} J.,  et~al., 2015, \mn@doi [\mnras] {10.1093/mnras/stu2058}, \href
  {http://adsabs.harvard.edu/abs/2015MNRAS.446..521S} {446, 521}

\bibitem[\protect\citeauthoryear{{Shaw}, {Weller}, {Ostriker}  \&
  {Bode}}{{Shaw} et~al.}{2006}]{2006_shaw}
{Shaw} L.~D.,  {Weller} J.,  {Ostriker} J.~P.,   {Bode} P.,  2006, \mn@doi
  [\apj] {10.1086/505016}, \href
  {http://adsabs.harvard.edu/abs/2006ApJ...646..815S} {646, 815}

\bibitem[\protect\citeauthoryear{Smethurst, Masters, Lintott  et~al.}{Smethurst
  et~al.}{2018}]{2018_smethurst}
Smethurst R.~J.,  Masters K.~L.,  Lintott C.~J.,   et~al., 2018, MNRAS, 473,
  2679

\bibitem[\protect\citeauthoryear{Springel}{Springel}{2005}]{2005_springel}
Springel V.,  2005, MNRAS, 364, 1105

\bibitem[\protect\citeauthoryear{Stewart, Maller, O{\~n}orbe  et~al.}{Stewart
  et~al.}{2017}]{2017_stewart}
Stewart K.~R.,  Maller A.~H.,  O{\~n}orbe J.,   et~al., 2017, ApJ, 843, 15

\bibitem[\protect\citeauthoryear{Teklu, Remus, Dolag  et~al.}{Teklu
  et~al.}{2015}]{2015_teklu}
Teklu A.~F.,  Remus R.-S.,  Dolag K.,   et~al., 2015, ApJ, 812, 24 pp.

\bibitem[\protect\citeauthoryear{Veale, Ma, Greene  et~al.}{Veale
  et~al.}{2017}]{2017_veale}
Veale M.,  Ma C.-P.,  Greene J.,   et~al., 2017, MNRAS, 471, 1428

\bibitem[\protect\citeauthoryear{Vogelsberger, Genel, Springel
  et~al.}{Vogelsberger et~al.}{2014}]{2014_vogelsberger}
Vogelsberger M.,  Genel S.,  Springel V.,   et~al., 2014, MNRAS, 444, 1518

\bibitem[\protect\citeauthoryear{Yurin \& Springel}{Yurin \&
  Springel}{2014}]{2014_yurin+springel}
Yurin D.,  Springel V.,  2014, MNRAS, 444, 62

\bibitem[\protect\citeauthoryear{Zavala, Frenk, Bower  et~al.}{Zavala
  et~al.}{2016}]{2016_zavala}
Zavala J.,  Frenk C.~S.,  Bower R.,   et~al., 2016, MNRAS, 460, 4466

\bibitem[\protect\citeauthoryear{Zjupa \& Springel}{Zjupa \&
  Springel}{2017}]{2017_zjupa}
Zjupa J.,  Springel V.,  2017, MNRAS, 466, 1625

\bibitem[\protect\citeauthoryear{de
  Vaucouleurs}{de~Vaucouleurs}{1959}]{1959_deVaucouleurs}
de Vaucouleurs G.,  1959, Handbuch der Physik,, 53, 275

\bibitem[\protect\citeauthoryear{de Zeeuw, Bureau, Emsellem  et~al.}{de~Zeeuw
  et~al.}{2002}]{2002_deZeeuw}
de Zeeuw P.~T.,  Bureau M.,  Emsellem E.,   et~al., 2002, MNRAS, 329, 513

\bibitem[\protect\citeauthoryear{van~de Sande, Bland-Hawthorn, Brough
  et~al.}{van~de Sande et~al.}{2017a}]{2017b_sande}
van~de Sande J.,  Bland-Hawthorn J.,  Brough S.,   et~al., 2017a, MNRAS, 472,
  1272

\bibitem[\protect\citeauthoryear{van~de Sande, Bland-Hawthorn, Fogarty
  et~al.}{van~de Sande et~al.}{2017b}]{2017a_sande}
van~de Sande J.,  Bland-Hawthorn J.,  Fogarty L.~M.,   et~al., 2017b, ApJ, 835,
  35pp

\makeatother
\end{thebibliography}




\appendix

\section{Testing simulation stability}

\subsection{The analytic potential}
\label{sec:appendix_anpot}
\begin{figure}
\centering
\includegraphics[width=\columnwidth]{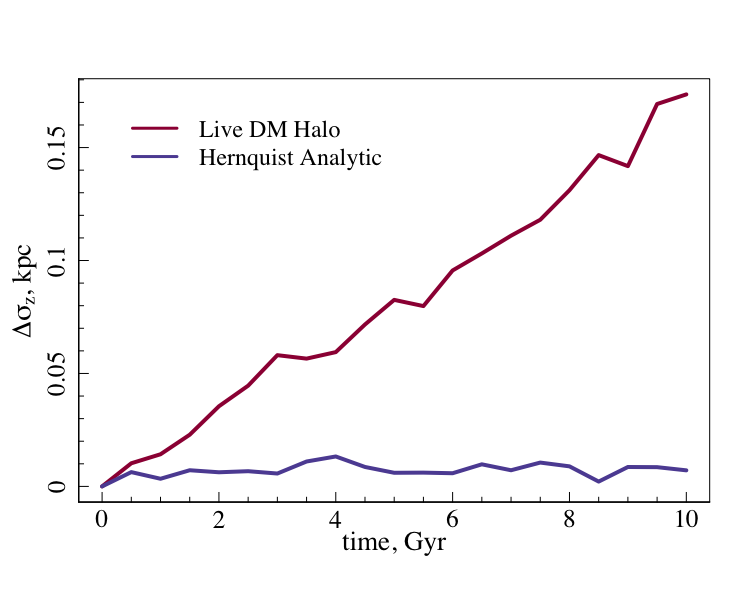}
\caption{Demonstrating the increasing standard deviation in the scale height density distribution of disk particles. The two plotted simulations contain $25,000$ disk particles representing a $\sim 10^{10}$ M$_{\odot}$ galaxy. The live DM halo represented by the red line contained $3\times10^6$ particles. Clearly, the scale height is not well maintained in this simulation when compared to the same disk in an analytic Hernquist potential.}
\label{fig:anpot1}
\end{figure}

When evolving the stellar components of the $N$-body models, we 
replace the underlying $N$-body dark matter halo with the analytical 
form assumed when setting up the initial conditions with \galic; we 
do this by modifying directly the \gadget\ source code so that the 
gravitational force and potential calculations already include the 
expected contribution from a dark matter halo (i.e. we use a static 
rather than a live dark matter halo). This is done for 
two reasons. The first is that it reduces the computational cost of our 
simulations by removing the need to simulate the time evolution of 
$\gtrsim 10^6$ dark matter particles when our interest is in the 
evolution of the stellar component. The second is implicitly related to
the first; particle discreteness gives rise to an unphysical increase in
the scale-height of the disc over time, as shown in Fig. 
\ref{fig:anpot1}, which modifies the velocity distribution of stars in 
the disc. The magnitude of this effect can be suppressed by
increasing the mass resolution of the dark matter halo relative to the 
stellar disk, but it is more expedient to adopt an analytical potential
instead, which allows us to simulate the $1\times10^6$ stars in the disk
needed for a careful measurement of kinematics at a relatively modest 
computational cost.

In practice, before evolving our initial conditions with \gadget, we 
remove all DM particles from the \galic\ output and ensure that
\gadget\ is initialised with the correct halo virial mass, concentration 
and virial overdensity, and spherically symmetric mass profile.
We use the modified Hernquist profile adopted by \galic\
\begin{equation}
\label{eq:hernq}
\rho_{dm}(r) = \frac{M_{dm}}{2 \pi} \frac{a}{r(r + a)^3} \;,
\end{equation}
where $M_{dm}$ is the total mass of the dark matter halo, $r$ is the radius at which the total enclosed mass is measured and $a$ is the scale factor of the halo as described by,
\begin{equation}
\label{eq:a}
a  = \frac{r_{200}}{c} \sqrt{2\left[ln(1+c) - c/(1+c)\right]} \;,
\end{equation}
in which $r_{200}$ is the virial radius of the halo and $c$ is the 
concentration of the DM halo. With these changes to \gadget\, we can
evolve our stellar-only initial conditions and ensure that they remain in
dynamical equilibrium.

\subsection{Testing the robustness of the analytic potential}
\begin{figure}
\centering
\includegraphics[width=\columnwidth]{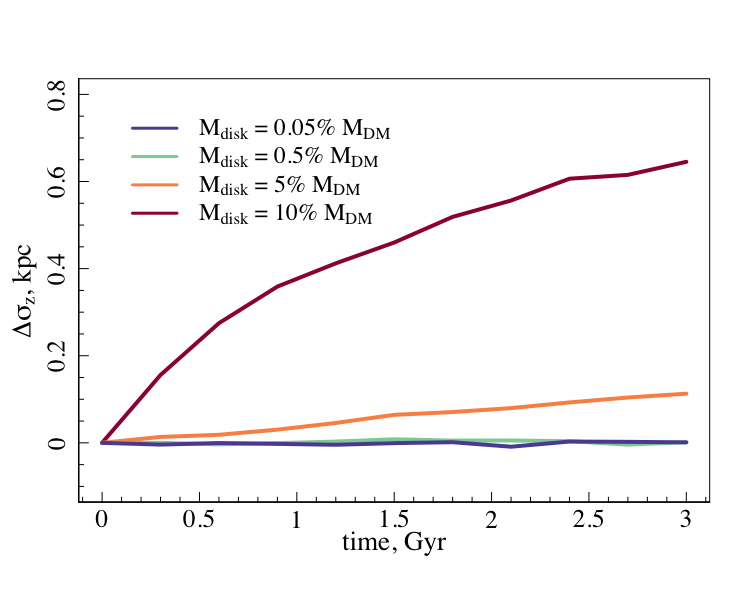}
\caption{Showing the change in standard deviation of the scale height 
density distribution of disk particles in four models with increasing disk 
mass. Each model has been evolved within the analytic DM potential. 
Clearly, as long as the mass of the disk stays below $5\%$ of the mass of 
the dark matter component, the scale height of the disk remains stable.}
\label{fig:anpot2}
\end{figure}

\begin{figure}
\centering
\includegraphics[width=\columnwidth]{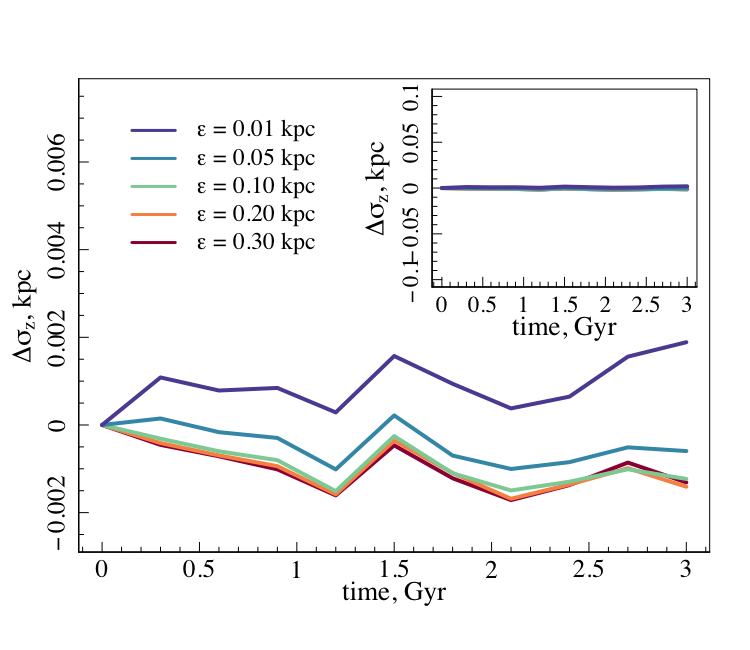}
\caption{Considering how the scale height density distributions of disk particles change over time in five models with different disk softening values. Each model has been evolved within the analytic DM potential. This implies that the simulation can be run at a range of resolutions without effecting the outcome of the kinematic analyses.}
\label{fig:anpot3}
\end{figure}

We have taken care to ensure that swapping the live dark matter halo for
its static, analytical, form does not affect the evolution of the stellar
disk. In Fig. \ref{fig:anpot2} and \ref{fig:anpot3}, we assess whether or 
not our choices of disk mass (Fig. \ref{fig:anpot2}) and gravitational softening (Fig. \ref{fig:anpot3}) influence the disk scale-height.

In Fig. \ref{fig:anpot2}, we check how disk mass impacts scale-height when 
the live halo is swapped for a static halo. \galic\ ensures that the 
initial conditions are in dynamical equilibrium, and it does this by 
modifying the dynamical structure of the dark matter halo in which the 
$N$-body stellar components are embedded; the more massive the disk, the 
larger the correction to the dynamical structure of the halo. Practically, 
this leads to a flattening of the dark matter halo, which is neglected 
when adopting the static halo in \gadget\ (which assumes spherical 
symmetry). Our tests suggest that stellar discs that contain more than 
$\sim 5\%$ of the mass of the DM component will start to deviate 
systematically from expected equilibrium evolution. As long as the 
catalogue consists of lower mass galaxies and only the kinematics of the 
stellar components are considered, the use of the analytic potential 
remains valid.

In Fig. \ref{fig:anpot3}, we show how the disk scale-height is influenced 
by gravitational softening. Softening leads to a modification of the 
gravitational potential within which particle trajectories are evolved, 
and so we expect to see differences only when the softenings are too 
large. Our results indicate that disk softening can be as large as 
$\epsilon = 0.3$ kpc (roughly $1/3^{\rm rd}$ of $z_0$) and as small 
as $0.01$ kpc without any adverse effects on the scale-height over time. 
This gives us confidence that our results are not compromised by choice of 
softening.

\section{Galactic angular momentum and \texorpdfstring{$\lambda_R$}{lr}}
\label{sec:appB}
In Appendix A of \citeauthor{2007_emsellem}'s \citeyear{2007_emsellem} 
paper, it's demonstrated how the observable spin parameter can be deduced 
from the Peebles definition, Eq.~\ref{eq:spin1}. We follow the spirit of 
this derivation and show how the observable spin parameter can be deduced 
from Bullock definition, Eq.~\ref{eq:spin2}. First, the equation can be rearranged,
\begin{align}
\lambda' &= \frac{J}{\sqrt{2} M V_c R} \;, \\
&= \frac{J}{\sqrt{2} MR  \sqrt{GM/R} } \;, \\
&= \frac{(J/M)}{\sqrt{2R} \sqrt{GM}} \;. \label{eq:sub}
\end{align}
This form is similar to equation A2 in appendix A of \cite{2007_emsellem}. Following their method of accounting for projection and mass-to-light conversions, we also use $\kappa_r$, $\kappa_J$, $\kappa_V$ and $\kappa_s$ to incorporate obervational effects on the radius, angular momentum, and the $V^2$ and $\sigma^2$ velocity moments.\\

This leaves us with the following expressions:
\begin{align}
J/M &= \kappa_J \left< R |V|\right> \;, \\
\sqrt{GM} &= \sqrt{\kappa_R \left< R (\kappa_V V^2 + \kappa_S \sigma^2) \right>} \;, \\
\sqrt{2R} &= \sqrt{2 \kappa_R \left<R\right>} \;,
\end{align}
which then substituting into Eq. \ref{eq:sub} results in a spin parameter,
\begin{align}
\lambda' &\sim \frac{\kappa_J \left<R|V|\right>}{\sqrt{2 \kappa_R \left<R\right>}\sqrt{\kappa_R \left< R (\kappa_V V^2 + \kappa_S \sigma^2) \right>}} \;, \\
&\sim \frac{\kappa_J}{\kappa_R \sqrt{2}} \frac{\left<R|V|\right>}{ \left< R \sqrt{\kappa_V V^2 + \kappa_S \sigma^2)} \right>} \;.
\end{align}
Finally, we use the second-order velocity moment $V^2 + \sigma^2$, where $\kappa_V = 1$ and $\kappa_S = 1$, such that we arrive at the same definition of $\lambda_R$ as \citeauthor{2007_emsellem}:
\begin{equation}
\lambda_R = \frac{\left<R|V|\right>}{ \left< R \sqrt{V^2 + \sigma^2)} \right>} \;.
\end{equation}
Substituting normal values for the angular momentum and radial conversion factors, $\kappa_J = 2$ and $\kappa_R = 3$ respectively, we arrive at $\lambda' \sim \sqrt{2} / 3 \, \lambda_R$. 


\bsp	
\label{lastpage}
\end{document}